\documentclass{elsart}
\usepackage{latexsym,amsmath,tabularx,amssymb,bm}
\usepackage{epsfig}

\long\def\comment#1{ }

\comment{
    \setlength{\textwidth}{16cm}
    \setlength{\textheight}{22.cm}
    \setlength{\oddsidemargin}{-0.04cm}
    \setlength{\evensidemargin}{-0.04cm}
    \setlength{\topmargin}{-1.31cm}
    }

\usepackage{graphicx}
\usepackage{amssymb}
\usepackage{makeidx}
\usepackage{graphicx}
\usepackage{multicol}

\newcommand{\nn}{\nonumber\\ }
\newcommand{\beq}{\begin{eqnarray}}
\newcommand{\eeq}{\end{eqnarray}}
\newcommand{\be}{\begin{eqnarray}}
\newcommand{\ee}{\end{eqnarray}}

    \newcommand{\abar}{\bar{\alpha}_s}
    \newcommand{\lan}{\langle}
    \newcommand{\ran}{\rangle}
\newcommand{\cal}{\mathcal}

\newcommand{\nt}{n^{(2)}}
\newcommand{\Tt}{T^{(2)}}

\def\simge{\mathrel{%
   \rlap{\raise 0.511ex \hbox{$>$}}{\lower 0.511ex \hbox{$\sim$}}}}
\def\simle{\mathrel{
   \rlap{\raise 0.511ex \hbox{$<$}}{\lower 0.511ex \hbox{$\sim$}}}}
\def\bigs{\mathrel{
   \rlap{\raise 0.531ex \hbox{$>$}}{\lower 0.531ex \hbox{$<$}}}}

\def\del{\partial}                              

\newcommand{\kk}{k_\perp}

\begin{document}
\begin{flushright}
~\vspace{-1.25cm}\\
{\small\sf SACLAY-T04/147}
\end{flushright}
\vspace{0.8cm}
\begin{frontmatter}

\title{A Langevin equation for high energy evolution with pomeron loops}

\author{E. Iancu\thanksref{th2}} and
\author{\ D. N. Triantafyllopoulos}
\address{Service de Physique Th\'eorique, CEA/DSM/SPhT,  Unit\'e de recherche
associ\'ee au CNRS (URA D2306), CE Saclay,
        F-91191 Gif-sur-Yvette, France}

\thanks[th2]{Membre du Centre National de la Recherche Scientifique
(CNRS), France.}

\date{\today}
\vspace{0.8cm}
\begin{abstract}
We show that the Balitsky--JIMWLK equations proposed to describe
non--linear evolution in QCD at high energy fail to include the
effects of fluctuations in the gluon number, and thus to correctly
describe both the low density regime and the approach towards
saturation. On the other hand, these fluctuations are correctly
encoded (in the limit where the number of colors is large) in
Mueller's color dipole picture, which however neglects saturation.
By combining the dipole picture at low density with the JIMWLK
evolution at high density, we construct a generalization of the
Balitsky hierarchy which includes the particle number fluctuations,
and thus the pomeron loops. After an additional coarse--graining in
impact parameter space, this hierarchy is shown to reduce to a
Langevin equation in the universality class of the stochastic
Fisher--Kolmogorov--Petrovsky--Piscounov (sFKPP) equation. This
equation implies that the non--linear effects in the evolution
become important already in the high momentum regime where the
average density is small, which signals the breakdown of the BFKL
approximation.

\end{abstract}

\end{frontmatter}
\newpage
\tableofcontents


\section{Introduction}
\setcounter{equation}{0}

Over the last decade, much progress has been realized towards
understanding the dynamics of QCD at high energies, in the vicinity
of the unitarity limit. New theoretical approaches have been
developed which encompass and extend the linear BFKL equation
\cite{BFKL} originally proposed to describe high energy evolution in
the leading--logarithmic approximation (with respect to $\ln s$), as
well as the earlier attempts to improve over the BFKL equation by
including unitarity corrections (or gluon saturation) in the form of
non--linear terms \cite{GLR,MQ85,BM87}.

In the mid nineties, Al Mueller \cite{AM94,AM95} has given an
elegant construction of the BFKL wavefunction of an energetic
hadron,  the `color dipole picture', which exploits the large--$N_c$
approximation (with $N_c$ the number of colors) to replace gluons by
color dipoles as the effective degrees of freedom at small $x$, and
gluon radiation by dipole splitting as the mechanism for evolution.
The average number of dipoles obeys BFKL equation, but the overall
description goes beyond the original BFKL approach by including the
many--body correlations generated through dipole splitting.
Accordingly, the evolution is expressed as a hierarchy of equations
for the dipole density correlations. The dipole picture cannot
describe interactions among the dipoles in the wavefunction (since
such interactions would lead to color configurations with higher
multipolar moments), so it cannot accommodate {\it gluon
saturation}. But it can be used to study the onset of unitarity
corrections in the elastic scattering between two sets of dipoles
which have evolved separately, provided the collision is viewed in
the center--of--mass frame and the total energy is not too large
\cite{AM95}. (The unitarization is brought in by the multiple
scattering of several pairs of dipoles from the two colliding
systems.) More generally, the dipole picture can be used to describe
the {\it dilute} (in the sense of non--saturated) part of a hadron
wavefunction up to arbitrarily high energies, and this is how we
shall actually use it throughout this paper.

A different formalism to study unitarization at high energy, in
which the unitarity corrections are encoded directly in the
evolution equations, has been developed by Balitsky \cite{B}. This
involves a hierarchy of equations describing the evolution of the
scattering amplitudes for the collision between a generic target and
a set of relatively simple `projectiles', which transform into each
other under the evolution. The target is not evolving, but is
generally taken to be a high--density system, or `color glass
condensate', represented by a strong color field. The evolution is
achieved by boosting the projectile, but in such a way that the
latter remains non--saturated, which implicitly restricts the
maximal energy allowed. Thus, the evolution of the projectile {\it
wavefunction} is in fact linear --- for a dipole projectile and in
the large--$N_c$ limit, this is precisely the dipole picture
discussed above\footnote{Recently, it has been explicitly shown, by
Levin and Lublinsky \cite{LL04}, that the large--$N_c$ version of
the Balitsky equations follows directly from the dipole picture for
the projectile, after allowing for multiple interactions with the
target.} ---, but Balitsky equations are nevertheless non--linear,
since written for the {\it scattering amplitudes} : the non--linear
effects correspond to multiple scattering between the components of
the projectile and the color field in the target.

Alternatively, one can keep the projectile unevolved, but use the
increase in the energy in order to boost the target, and then study
the small--$x$ evolution of the strong color fields there. In doing
so, some of the non--linear effects which were interpreted as
multiple scattering from the perspective of projectile evolution
will now appear as saturation effects in the evolution of the
target. But to study this, one needs a formalism capable to deal
with non--linear effects in the evolution of the wavefunction. For
the reasons explained before, this cannot be a large--$N_c$
formalism in terms of dipoles, but rather it must allow for all the
possible color multipoles, and the simplest and most natural way to
do so is to work in terms of gluons. So far, the only formalism of
this type is the {\it color glass condensate} (CGC)
\cite{MV,K96,JKMW97,JKLW97,RGE,SAT} (for reviews see Ref.
\cite{CGCreviews}), in which the small--$x$ gluons are described as
classical color fields generated by color sources at larger values
of $x$, which are randomly distributed (in color and coordinate
space) with a weight function which evolves with the energy. This
evolution is governed by a functional Fokker--Planck equation for
the weight function, the JIMWLK equation \cite{JKLW97,RGE,W}, which
is equivalent to an infinite hierarchy of equations for the
correlation functions of the classical color fields. When applied to
the scattering between a simple projectile and the CGC, the JIMWLK
evolution reproduces Balitsky equations for the scattering
amplitudes. In what follows, we shall refer to these equations as
the Balitsky--JIMWLK equations.

The Balitsky--JIMWLK equations will be further discussed in Sect.
\ref{BJIMWLK} below, but here we would like to emphasize a few
important points: {\it i)} The evolution described by these
equations is {\it stochastic} : Through the {\it non--linear terms}
present in these equations, $n$--point functions with different
values of $n$ will mix under the evolution, thus generating new
correlations with increasing energy. {\it ii)} The non--linear terms
are associated with the presence of {\it strong classical fields}
(the field of the target in the approach by Balitsky, or the field
created by color sources at higher values of $x$ in the CGC).
Therefore, in the dilute regime where the fields are weak, the
equations can be linearized, and once we do so, they decouple from
each other (so that the various $n$--point functions obey
independently the BFKL evolution). This should be contrasted to the
dipole picture, where correlations beyond the BFKL equation appear
already in the dilute regime. {\it iii)} Even in the full equations
with the non--linear terms included, the stochastic aspects turn out
to become inessential at large\footnote{By ``large--$N_c$"  we
understand here the {\it high--energy} version of the large--$N_c$
approximation, due to Mueller \cite{AM94,AM95}, in which gluon
exchanges which are suppressed by factors of $1/N_c^2$ are still
counted as leading order effects provided they are enhanced by
appropriate powers of the energy. The neglected terms are those
which, for a given power of the energy, are suppressed by higher
powers of $1/N_c^2$ than the leading--order terms.} $N_c$. For
instance, in the absence of correlations in the initial conditions,
the large--$N_c$ version of the Balitsky hierarchy boils down to a
single, non--linear, equation, which is deterministic. This is the
equation originally derived by Kovchegov \cite{K} and generally
referred to as the Balitksy--Kovchegov (BK) equation. This suggests
that the stochastic aspects of the Balitsky--JIMWLK equations are to
be attributed to {\it color fluctuations} in the high--density
regime, which are suppressed at large $N_c$, unlike the colorless
{\it fluctuations in the particle number}, as encoded in the dipole
picture, which are present already in the dilute regime and for
large $N_c$.

The previous discussion --- which will be further substantiated and
illustrated with a few Feynman diagrams in Sect. \ref{BJIMWLK} ---
points out towards an insufficiency of the Balitsky--JIMWLK
equations, which fail to include the correlations associated with
fluctuations in the particle number.
Because of that, the mean field approximation (the BK equation)
should work very well for these equations at large $N_c$ and for
uncorrelated initial conditions (like the scattering off a large
nucleus), and this is indeed what is seen in numerical studies of
JIMWLK evolution \cite{RW03}. So far, most studies of saturation and
unitarity have in fact focused on the BK equation, which is much
easier to handle in practice than the general equations, and whose
properties are by now rather well understood
\cite{LT99,AB01,Motyka,LL01,GBS03,Nestor03,Nestor04,SCALING,MT02,DT02,MP03}.

However, there were indications already in the dipole picture (see
especially the numerical studies by Salam \cite{Salam95,AMSalam96})
that the effects of fluctuations should be important, in particular,
in the evolution from a low density regime to a high density one,
and thus in the approach towards saturation. Indeed, in the dilute
regime, the particle number fluctuations are a natural mechanism to
generate higher--point density correlations, which then evolve into
the non--linear terms responsible for saturation. E.g., if one
starts with an isolated gluon (or dipole) at high transverse
momentum ($\kk$), as radiated from the bulk of particles at lower
$\kk$, then higher--point correlations at high $\kk$ --- which were
originally absent --- will get built in the early stages of the
evolution, as correlations in the splitting of the original dipole
or its close descendence. Later on, these correlations get amplified
by the standard BFKL evolution, and eventually influence the
evolution of the lower--point correlations when the density is high
enough.

The importance of fluctuations in the evolution towards saturation
has been reiterated in recent publications \cite{IM032,MS04,IMM04}.
In Ref. \cite{IM032} it has been argued that the fluctuations slow
down the approach towards the unitarity limit as compared to the
mean field approximation (MFA) \cite{SAT,LT99}. In Ref. \cite{MS04},
a modified MFA has been proposed, in which the unitarity constraints
associated with fluctuations have been simulated by imposing a
barrier at high momenta on the BFKL equation. The solution to the
ensuing equation shows that fluctuations reduce considerably the
rate of growth of the saturation momentum with the energy (or
`saturation exponent'), and that the corresponding mean--field limit
is reached only very slowly, {\it logarithmically}, as $\alpha_s\to
0$. Finally, in Ref. \cite{IMM04}, the event--by--event picture of
the evolution has been considered, in which fluctuations appear
naturally because, in a given event, the particle occupation numbers
are {\it discrete}. Based on this picture, a correspondence has been
established between high--energy evolution in QCD and a class of
stochastic particle models which are actively studied in statistical
physics. By using known results for the latter, the authors of Ref.
\cite{IMM04} have confirmed the result in Ref. \cite{MS04} about the
slow convergence of the saturation exponent to its mean--field
value, and further argued that fluctuations should eventually wash
out the {\it geometric scaling} \cite{geometric} property of the
average amplitude, which was known to hold for the BK equation
\cite{SCALING,MT02,DT02,MP03}. However, the arguments in Ref.
\cite{IMM04} cannot predict {\it how fast} (i.e., after what energy
evolution) is geometric scaling violated if one starts with an
initial condition that shows scaling, nor they allow one to study
the {\it preasymptotic} evolution at intermediate energies and for
realistic values of $\alpha_s$.

More importantly, there was some confusion in Ref. \cite{IMM04}
about what should be the right formalism to study this evolution.
Namely, it has been assumed there that the correct evolution law is
the one provided by the Balitsky--JIMWLK equations, but this cannot
be right, since these equations miss the relevant fluctuations, as
we have argued above. It is then natural to ask, what is/are the
equation(s) that one should solve (e.g., numerically) in order to
study the interplay between fluctuations and saturation, and thus be
able to verify the results in Refs. \cite{MS04,IMM04} and improve
over them. In other terms, what is the correct formalism to describe
evolution in QCD at high energy ?

It is the main purpose of this paper to provide an answer to this
question, in the form of a new set of evolution equations which
generalize the (large--$N_c$ version of the) Balitsky--JIMWLK
equations by approximately including the effects of particle number
fluctuations. Our basic observation is that, in fact, we do have the
necessary ingredients to describe both fluctuations and saturation
--- these are the color dipole picture, and the JIMWLK evolution,
respectively --- and that, although these ingredients may look too
different from each other to be simply married in a unified
theoretical description of the lightcone {\it wavefunction}, they
can still be naturally merged with each other in the evolution
equations for {\it scattering amplitudes}. Since relying on the
dipole picture, our subsequent construction is limited to the
large--$N_c$ limit, to which we shall restrict ourselves throughout
the following analysis. It remains as an open problem at this stage
whether it would be possible to develop a wavefunction formalism
which incorporates all these physical ingredients (e.g., through a
suitable extension of the JIMWLK formalism), and thus rederive the
results for scattering amplitudes that we shall present here, together
with their generalization to arbitrary $N_c$.

Let us briefly anticipate here the main steps in these construction,
and explain the other approximations that we shall need. After briefly
explaining the physical picture of particle number fluctuations in the
dipole picture in Sect. \ref{General}, we shall then construct,
in Sect. \ref{Dipoles}, the equations which describe the evolution
of the many--body density correlations in this formalism.
To that aim, we shall use the recent formulation of the dipole
picture in Refs. \cite{IM031,LL03}, which is more convenient for a
study of fluctuations since it follows explicitly the evolution of
a given configuration of $N$ dipoles. The equations that we shall
obtain in Sect. \ref{Dipoles} have been already presented by Levin and
Lublinsky in a very recent publication \cite{LL04}, but our respective
derivation will be somewhat different, and also better suited for our present
purposes.

In the meantime, in Sect. \ref{Toy}, we shall also analyze a simple
stochastic particle model borrowed from statistical mechanics (a
zero--dimensional version of the model in Ref. \cite{PL99}), which
in spite of its formal simplicity has the conceptual advantage over
the dipole picture to include both fluctuations and recombination
(and thus saturation). The explicit manipulations permitted by the
simplicity of this model will help us developping some physical
intuition, and will also serve as a guidance for the corresponding
manipulations in QCD, where such a unified theoretical description
of the wavefunction
is still lacking. A more elaborated, one--dimensional, version of the
same model (the one which is actually considered in Ref. \cite{PL99})
will be then discussed in the Appendix. This will give us the
opportunity to introduce the stochastic
Fisher--Kolmogorov--Petrovsky--Piscounov (sFKPP) equation, which
in this context emerges as an {\it exact} equation (mathematically
equivalent to the original model), and which will also emerge
as an approximation to the corresponding equations in QCD.

In Sect. \ref{Langevin} we construct the new equations for
scattering amplitudes in QCD, which are the main results of this
paper. First, in Sect. 6.1, we relate dipole correlations in the
target to scattering amplitudes for projectile dipoles, by using
approximations which make sense in the dilute regime. This allows us
to translate, in Sect. 6.2, the equations for densities constructed
in Sect. \ref{Dipoles} into corresponding equations for the
scattering amplitudes. These equations are still linear, as
appropriate in the low--density regime, but they include the effects
of dipole number fluctuations in the target (so they form already a
non--trivial hierarchy). Then, these equations are extended to the
high--density regime by adding the non--linear terms expected from
the JIMWLK evolution (i.e., the same non--linear terms as in the
Balitsky hierarchy). Because of the combined effect of fluctuations
(`gluon splitting') and recombination  (`gluon merging'), the final
equations generate pomeron loops through iterations. In Sect. 6.3,
we show that, after a coarse--graining in the impact parameter
space, the whole hierarchy can be equivalently replaced by a
specific Langevin equation, which is formally the BK equation
supplemented with a multiplicative noise term.

In our last section \ref{Physics}, we explore some physical and
mathematical consequences of the new equations. In Sect. 7.1, we
show that, mediating a gradient expansion of the BFKL kernel known
as the `diffusion approximation', the Langevin equation of Sect.
\ref{Langevin} can be cast into the form of the sFKPP equation. The
latter emerges as an effective equation for a variety of problems in
physics, biology, and chemistry (see Refs. \cite{Saar,Panja} for
recent reviews), and it has been extensively studied in recent years
in the statistical physics literature. By using some known results
about this equation, we confirm (in Sect. 7.2) previous findings in
Refs. \cite{MS04,IMM04}, and thus conclude that the present Langevin
equation is indeed the evolution law underlying the physical
discussion in Ref. \cite{IMM04}. Finally, in Sect. 7.3, we discuss a
rather dramatic consequence of the stochastic nature of the
evolution, which entails the breakdown of the BFKL approximation in
the high momentum regime, where the average gluon density, or the
average scattering amplitude, are small. Because of fluctuations,
the saturation momentum becomes a random quantity which can take
different values from one event to another, or from one impact
parameter to another (for the same event). Then the evolution
develops high--density `spots' where the gluons are at saturation
even for relatively high transverse momenta, well above the {\it
average} saturation momentum. At sufficiently high energy, all the
correlations are dominated by such dense spots, and the mean field
approximation (i.e., BFKL equation at high $\kk$) breaks down. This
may look unexpected, but in fact it can be related to a similar
behavior observed by Salam in his Monte--Carlo studies of the dipole
picture \cite{Salam95,AMSalam96}. The `dense spots' were clearly
seen in those numerical simulations, but of course in that context
the local density was never saturating.

\section{Physical motivation}
\label{General} \setcounter{equation}{0}

Given a high energy hadron with rapidity $Y$, we would like to
understand fluctuations in the tail of the gluon distribution at
transverse momenta $\kk\gg Q_s(Y)$ (with $Q_s(Y)$ the saturation
momentum), and, in particular, the influence that such fluctuations
may have
--- through their subsequent evolution --- on the approach towards
saturation (or towards unitarity in the collision with an external
projectile). The high--$\kk$ tail is a priori a dilute regime in
which the gluon occupation numbers are small, of ${\cal O}(1)$, so
we expect important fluctuations associated with the discreteness of
the particle (here, gluon) number. Given that the gluon number
density is represented, in a field theoretical formulation, by the
two--point correlation function of the color fields in the hadron
wavefunction, we deduce that the relevant fluctuations should be
encoded in the four--point, and higher, correlations. Alternatively,
and simpler, in the large--$N_c$ limit to which we shall stick
throughout this paper, the tail of the distribution can be described
in Mueller's dipole picture \cite{AM94,AM95}, in which gluons are
effectively replaced by quark--antiquark pairs of zero transverse
size and in a color octet state. The `color dipoles' then emerge as
color singlet states built with the quark component of some gluon
and the antiquark component of some other gluon with roughly the
same rapidity. In this picture, which applies so long as the gluon
density is low enough for saturation effects to be negligible, the
(unintegrated) gluon distribution is measured by the average dipole
number density $\langle n(r,Y)\rangle$ which obeys BFKL equation
\cite{BFKL}. But the dipole picture goes beyond the strict BFKL
equation by including {\it correlations} in the distribution of
dipoles, namely those correlations which follow from dipole
splitting in the course of the evolution. The simplest such
correlations, which also encompass the particle number fluctuations
that we are mainly interested in, are those encoded in the pairwise
dipole number density $\langle \nt(r_1,r_2,Y)\rangle$ which is a
measure of the probability to find simultaneously two dipoles, with
transverse sizes $r_1$ and $r_2$, respectively (see Sect.
\ref{Dipoles} below for a more precise definition).

Since it is preferable to consider quantities which are measurable
(at least, in principle), we shall study the effect of fluctuations
on the scattering between the dilute hadronic target (described
within the dipole picture) and an external projectile, which is
itself chosen as a set of dipoles in some fixed configuration. This
choice is convenient since, as well known,
the dipole--dipole scattering is {\it
quasi--local} in phase--space: a dipole projectile essentially
counts the numbers of dipoles in the target having the same
transverse size\footnote{More precisely, (quasi)locality in the
dipole size holds so long as the dipole distribution in the target
is in a genuine BFKL regime, i.e., it is characterized by some
`anomalous dimension' ; see Sect. \ref{Langevin} for details.} and
impact parameter as itself. Specifically, the scattering amplitude
for a single dipole can be estimated as $T(r,b,Y)\sim \alpha_s^2
f(r,b,Y)$, where $r$ and $b$ denote respectively the size and the
impact parameter of the projectile dipole, $\alpha_s^2$ measures the
scattering amplitude for two dipoles with roughly the same size and
nearby impact parameters, and $f(r,b,Y)$ is the (dimensionless)
dipole occupation factor in the target, and is related to the
corresponding number density via $f(r,b,Y)\sim r^4 n(r,b,Y)$ (see
Sect. \ref{Langevin}).

To be sensitive to fluctuations, the projectile must involve at
least two dipoles, and here we shall consider the case where it
contains exactly two. Both dipoles are assumed to be small, so the
individual scattering amplitudes $\langle T_i\rangle\sim \alpha_s^2
\langle f_i\rangle$ (with $\langle T_i\rangle\equiv\langle
T(r_i,b_i,Y)\rangle$ and $i=1,2$) are both small: $\langle
T_i\rangle\ll 1$. The question is then, how large can be $\langle
\Tt\rangle\sim \alpha_s^4 \langle f_1f_2\rangle$ (the scattering
amplitude for the simultaneous scattering of both dipoles) ?
Clearly, in the absence of correlations in the target, or if the
correlations are only weak, one has $\langle f_1f_2\rangle\simeq
\langle f_1\rangle \langle f_2\rangle$, and then $\langle
\Tt\rangle\simeq \langle T_1 \rangle \langle T_2 \rangle$ is much
smaller than either $\langle T_1 \rangle$ or $\langle T_2 \rangle$.
However, as we discuss now, there are kinematical situations in
which one expects strong correlations among the dipoles in the
target, which will drastically enhance $\langle \Tt\rangle$ with
respect to its estimate $\langle T_1 \rangle \langle T_2 \rangle$
for incoherent scattering.

The first such a situation is not really relevant for the main
stream of this paper, and the only reason for mentioning it here is
to distinguish it from the more interesting case to be discussed
after. Namely, this is the situation in which both external dipoles
scatter off the {\it same} dipole in the target. This is
advantageous in the {\it very} dilute regime where $\langle f\rangle
\ll 1$ (or $\langle T\rangle \ll \alpha_s^2$), since for such a
process there is a single low--density penalty factor, and the
corresponding contribution $\langle \Tt\rangle\sim \alpha_s^4
\langle f \rangle$ is indeed much larger than the uncorrelated piece
$\sim \alpha_s^4 \langle f \rangle^2$. On the other hand, in view of
the previous considerations on the quasi--locality of dipole--dipole
scattering, it is clear that this situation requires a rather fine
tuning between both the transverse sizes and the impact parameters
of the two incoming dipoles ($r_1\sim r_2$ and $b_1 \sim b_2$),
which as we shall see is not an interesting configuration for the
evolution of the gluon distribution to high energies.

The second situation, which is more relevant for our purposes here,
is the one in which the two dipoles making the projectile are {\it
contiguous} in transverse space, meaning that the quark leg of one
dipole is close to the antiquark leg of the other. If we use ${\bm
x}_1$ and ${\bm y}_1$ (${\bm x}_2$ and ${\bm y}_2$) to denote the
transverse positions of the quark and respectively antiquark leg of
the first (second) dipole, then contiguous configurations correspond
to either ${\bm y}_1\approx {\bm x}_2$ or to ${\bm x}_1\approx {\bm
y}_2$. Such configurations are interesting because the scattering
amplitude $\langle \Tt\rangle$ for two contiguous dipoles enters the
evolution equation for the scattering amplitude of a single dipole
in the regime where unitarity corrections become important (see
Sect. \ref{BJIMWLK}). The reason why $\langle \Tt\rangle$ is
enhanced as compared to $\langle T\rangle^2$ for such configurations
has to do with the dynamics of the evolution in the dipole picture:
When increasing rapidity from $Y$ to $Y+dY$, a dipole with legs at
${\bm x}$ and ${\bm y}$ can evolve by radiating one soft (i.e.,
small--$x$) gluon located at ${\bm z}$, a process which at
large--$N_c$ is tantamount to the original dipole (${\bm x},\,{\bm
y}$) splitting into two new dipoles (${\bm x},\,{\bm z}$) and (${\bm
z},\,{\bm y}$), which are contiguous. This evolution leads to an
{\it increase} in the scattering amplitude of a projectile made of
two contiguous dipoles (with appropriate transverse coordinates)
which is proportional to the average density $\langle n({\bm
x},\,{\bm y})\rangle_Y$ of the parent dipole at rapidity Y. That is,
it is the dipole density $\langle n\rangle$, rather than the dipole
{\it pair} density $\langle \nt\rangle$, which acts directly as a
source for $\langle \Tt\rangle$, and this is certainly advantageous
in the dilute regime where\footnote{More correctly, this inequality
should be written for the dimensionless occupation numbers, as
$\langle f f\rangle \ll \langle f \rangle$.} $\langle \nt \rangle
\ll \langle n \rangle$.

For instance, if one starts with a single dipole (${\bm x}_0,\,{\bm
y}_0$) at $Y=0$, then the only way to find a pair of dipoles after
an evolution $dY$ is that this pair be made of the dipoles (${\bm
x}_0,\,{\bm z}$) and (${\bm z},\,{\bm y}_0$), with arbitrary ${\bm
z}$. More generally, the $Y$--evolution of an arbitrary target will
generate ``high--$k_\perp$ fluctuations" (i.e., small dipoles which
split off the larger, preexisting dipoles) with low occupation
numbers, of ${\mathcal O}(1)$, and the only way to increase the
dipole occupancy in those high--$k_\perp$ bins in the next few steps
of the evolution is through the splitting of the original
fluctuations. It is only when the average pair density $\langle
\nt\rangle$ becomes large enough that the normal BFKL evolution
(here, for $\langle \nt\rangle$) takes over, and the fluctuations
(at that particular value of $k_\perp$) cease to play a role. We
see that a high--$k_\perp$ fluctuation play the crucial role of a
{\it seed} for extending the gluon distribution towards larger
transverse momenta with increasing $Y$.

\begin{figure}[t]
  \centerline{\epsfxsize=5cm\epsfbox{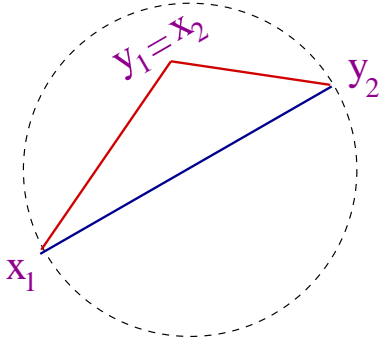}}
    \caption{\sl The geometry of dipole splitting}
\end{figure}

It is in fact easy to turn some of the above considerations into
more explicit formulae: The differential probability for a dipole
(${\bm x},\,{\bm y}$) to split into new dipoles (${\bm x},\,{\bm
z}$) and (${\bm z},\,{\bm y}$) is known within perturbative QCD (to
leading-log accuracy with respect to $Y=\ln 1/x$) as
($\bar\alpha_s\equiv \alpha_s N_c/\pi$)
 \be\label{dipolesplit}
d{\mathcal P}(\bm{x},\bm{y}|\bm{z})\,=\,
\frac{\bar\alpha_s}{2\pi}\,\frac{(\bm{x}-\bm{y})^2}{(\bm{x}-\bm{z})^2
(\bm{z}-\bm{y})^2}\,{d^2{\bm z}}\,dY\,.
 \ee
This formula, together with elementary geometry considerations (see
Fig. 1), immediately imply (with ${\bm r}_1={\bm x}_1 -{\bm y}_1$
and ${\bm r}_2={\bm x}_2 -{\bm y}_2$) :
 \be\label{dn20}
\hspace*{-0.6cm} \frac{\partial}{\partial Y}\,\langle \nt({\bm
x}_1,{\bm y}_1 \,;{\bm x}_2, {\bm y}_2)\rangle_Y\Big |_{\rm fluct.}
\!&=& \frac{\bar\alpha_s}{2\pi}\,\frac{({\bm r}_1+{\bm r}_2)^2}{{\bm
r}_1^2 {\bm r}_2^2}\big\{ \delta^{(2)}({\bm x}_2-{\bm y}_1)\langle
n({\bm x}_1,\,{\bm y}_2)\rangle_Y \nn &{}&\qquad\qquad +\,
\delta^{(2)}({\bm y}_2-{\bm x}_1)\langle n({\bm x}_2,\,{\bm
y}_1)\rangle_Y\big\}\,.\,\,
 \ee
As indicated in the l.h.s. of Eq.~(\ref{dn20}), this is just the
``fluctuating" contribution to the evolution of $\langle
\nt\rangle$, in which the two measured dipoles are generated via the
splitting of a unique original dipole. The general equation for
$\partial\langle \nt\rangle/ {\partial Y}$, to be constructed in
Sect. \ref{Dipoles} within the dipole picture, involves also terms
linear in $\langle \nt\rangle$, which correspond to the usual BFKL
evolution. By also using the
relation $T({\bm r}, {\bm b},Y)
\sim \alpha_s^2 \,r^4 n({\bm r},{\bm b},Y)$,
between (target) dipole densities and (projectile) dipole
scattering amplitudes, it is possible to
transform evolution equations for the dipole densities, so like
Eq.~(\ref{dn20}), into corresponding equations for the scattering
amplitudes. This will be discussed in Sect. \ref{Langevin}.

\section{The Balitsky--JIMWLK equations and beyond}
\label{BJIMWLK} \setcounter{equation}{0}

Our discussion of the Balitsky--JIMWLK equations in this section
will be rather schematic, as our intention is merely to show --- by
inspection of the structure of these equations, and with the help of
a few Feynman diagrams --- that these equations do not include the
effects of gluon number fluctuations in the target wavefunction.
Since, as we shall later argue, these fluctuations serve as the
initiators of the ``pomeron loops" in the target, our conclusion
also implies that the evolution generated by these equations fails
to include the pomeron loops. This failure was already recognized in
the literature in relation with the Kovchegov equation, which is a
mean field approximation to the Balitsky--JIMWLK equations, but to
our knowledge it has not been argued so far for the general
equations.

For consistency for the other developments in this paper, we shall
restrict also the present discussion to the (high--energy version of
the) large--$N_c$ limit, in which Balitsky--JIMWLK equations close
in the space of dipoles. That is, they reduce to a hierarchy of
equations for the evolution of the scattering amplitudes of a set of
$N$ dipoles, with $N=1,2,\dots$, which scatter off a generic target.
The color fields in the target can be strong
--- that is, the target can be at saturation (a `color glass
condensate') ---, but the equations include non--linear effects
which ensure that scattering is unitary. As we shall see, the
non--linear effects in the evolution can be interpreted as either
multiple scattering, or saturation, depending upon the perspective
from which one views the evolution (as projectile or, respectively,
target evolution).

The simplest way to present the (dipolar version of the)
Balitsky--JIMWLK equations is to notice that the whole hierarchy can
be generated from the following ``operator" equation:
 \be\label{Balitsky}
{\partial \, T_Y({\bm x},{\bm y})\over {\partial Y}}&=&
\frac{\bar\alpha_s}{2\pi}\int d^2{\bm z} {{({\bm x}-{\bm y})^2}
\over {({\bm x}-{\bm z})^2 ({\bm y}-{\bm z})^2}}\\ &{}& \quad
\big\{-T_Y({\bm x},{\bm y})+T_Y({\bm x},{\bm z})+T_Y({\bm z},{\bm
y})-T_Y({\bm x},{\bm z})T_Y({\bm z},{\bm y})\big\}.\nonumber \ee
 By ``operator equation" we simply mean that, in order to deduce
the equation satisfied by the $N$--point function $\langle
T^{(N)}\rangle_Y\equiv \langle T(1)T(2)\cdots T(N)\rangle_Y$ (with
$T(i)\equiv T({\bm x}_i,{\bm y}_i)$), it is sufficient to multiply
Eq.~(\ref{Balitsky}) by appropriate powers of $T$ and then use
Leibniz' rule; e.g. :
 \be\label{Balitsky2T} {\partial\over
\partial Y}\, \big\langle T(1)T(2)\big\rangle_Y =
\big\langle{\partial T(1)\over \partial Y}\,T(2)\big\rangle_Y +
\big\langle T(1)\,{\partial T(2)\over\partial Y}\big\rangle_Y\,.\ee
In particular, the equation obeyed by the average scattering
amplitude of a single dipole is immediately obtained as:
 \be\label{EQT1}
{\partial \,\over {\partial Y}}\, \langle T({\bm x},{\bm y})
\rangle_Y&=& \frac{\bar\alpha_s}{2\pi}\int d^2{\bm z} {{({\bm
x}-{\bm y})^2} \over {({\bm x}-{\bm z})^2 ({\bm y}-{\bm z})^2}}\\
&{}& \quad \big\langle-T({\bm x},{\bm y})+T({\bm x},{\bm z})+T({\bm
z},{\bm y})-T({\bm x},{\bm z})T({\bm z},{\bm
y})\big\rangle_Y.\nonumber
 \ee
As anticipated in Sect. \ref{General}, the r.h.s. of this equation
involves the scattering amplitude $\langle \Tt\rangle=\langle T({\bm
x},{\bm z})T({\bm z},{\bm y})\rangle$ for two contiguous dipoles.
More generally, the equation for $\langle T^{(N)}\rangle_Y$ involves
also $\langle T^{(N+1)}\rangle_Y$, so Eq.~(\ref{Balitsky}) generates
an infinite hierarchy of equations which decouple from each other
only in the weak scattering regime, where $\langle T^{(N+1)}\rangle
\ll \langle T^{(N)}\rangle$ and each $N$--point function obeys
separately the BFKL equation for evolution in any of its $N$
arguments.

But the hierarchy generated by Eq.~(\ref{Balitsky}) is sufficiently
simple for the corresponding evolution to be {\it
quasi--deterministic}. It is indeed easy to check that, if the
initial conditions at $Y=Y_0$ are chosen in {\it factorized} form,
i.e., $ \langle T(1)\cdots T(N)\rangle_0 = \langle T(1)\rangle_0
\cdots \langle T(N)\rangle_0 $, then this factorized form is
preserved by the evolution up to arbitrarily large $Y$ :
schematically, $\langle T^{(N)}\rangle_Y = \langle T\rangle_Y^N$
with the one--point function $\langle T\rangle_Y$ obeying the BK
equation (i.e., the equation obtained by replacing $T\to \langle
T\rangle_Y$ into Eq.~(\ref{Balitsky})). More generally, it has been
shown in Refs. \cite{JP04,Janik} that the hierarchy generated by
Eq.~(\ref{Balitsky}) admits a one--parameter family of fully
factorized exact solutions. This strongly suggests that, in their
simplified form valid at large $N_c$, the Balitsky--JIMWLK equations
do not generate new correlations, but only propagate those already
encoded in the initial conditions. This simplifying feature does not
hold also for the {\it full} equations, which include additional,
multipolar, operators. But the correlations induced by these
operators are suppressed  by powers of $1/N_c^2$, and thus cannot be
associated with fluctuations in the gluon (or dipole) number.
Rather, as discussed in the Introduction, they describe {\it color}
fluctuations.

 This conclusion is further substantiated
by an analysis of the diagrammatic content of the Balitsky--JIMWLK
equations, to which we now turn. At this point one should recall
that the structure of the perturbation theory, and thus the form of
the diagrams, depends upon the frame in which one is viewing the
evolution:

{\bf a)} In the original derivation by Balitsky \cite{B}, the
evolution is implemented by boosting the {\it projectile}, which
then evolves through (small--$x$) gluon radiation. In the
large--$N_c$ limit in which the projectile is a collection of $N$
dipoles, its evolution amounts to the splitting of one of these
dipoles into two new dipoles, followed by the interaction between
the final system of $N+1$ dipoles and the target. Then
Eq.~(\ref{Balitsky}) applies to one such a dipole which has split,
and the various terms there are easily interpreted: The two linear
terms with positive sign, $T_Y({\bm x},{\bm z})$ and $T_Y({\bm
z},{\bm y})$, describe the independent scattering of the daughter
dipoles with the target, the quadratic term with a negative sign
corrects for an overcounting of their simultaneous scattering, and
the linear term with a negative sign is the ``virtual term" which
expresses the possibility that the parent dipole $({\bm x},{\bm y})$
survive without splitting.

In this picture, the non--linear terms in the evolution are thus
associated with multiple scattering, but they indirectly reflect
saturation effects in the {\it target}. But as far as the {\it
projectile} is concerned, there is still no saturation: The $N$
dipoles composing the projectile are not allowed to interact with
each other, which is a good approximation only so long as
$\alpha_s^2 N\ll 1$. Since the total number of dipoles within the
projectile grows exponentially with $Y$, this is clearly a low
energy approximation.

\begin{figure}[t]
    \centerline{\epsfxsize=5cm\epsfbox{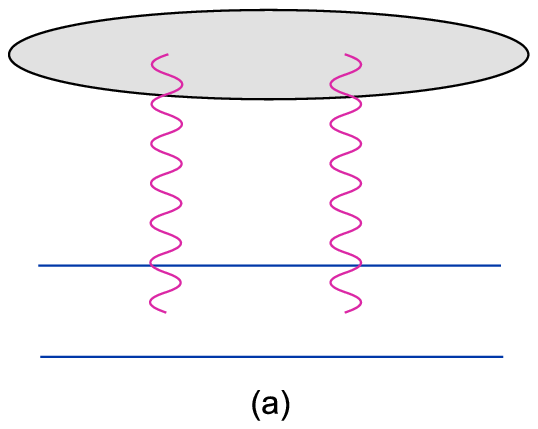}
    \hspace{0.5cm}
    \epsfxsize=5cm\epsfbox{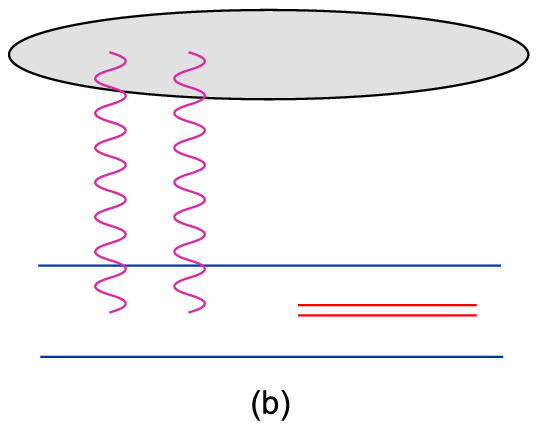}}
    \vspace{1cm}
    \centerline{\epsfxsize=5cm\epsfbox{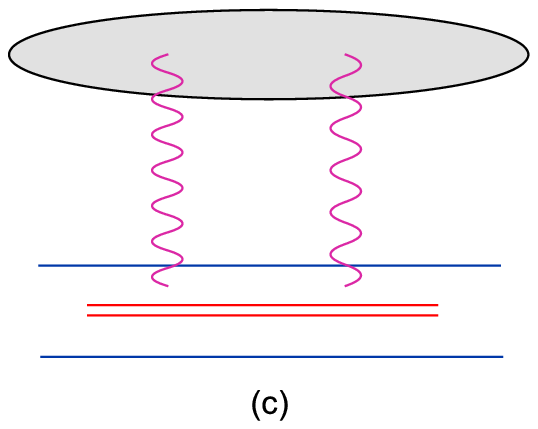}
    \hspace{0.5cm}
    \epsfxsize=5cm\epsfbox{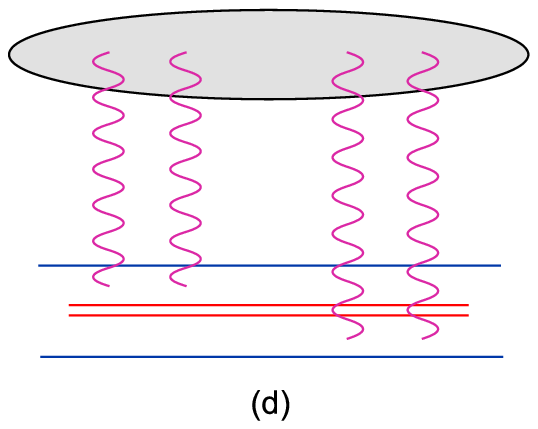}}
    \vspace{1cm}
    \centerline{\epsfxsize=5cm\epsfbox{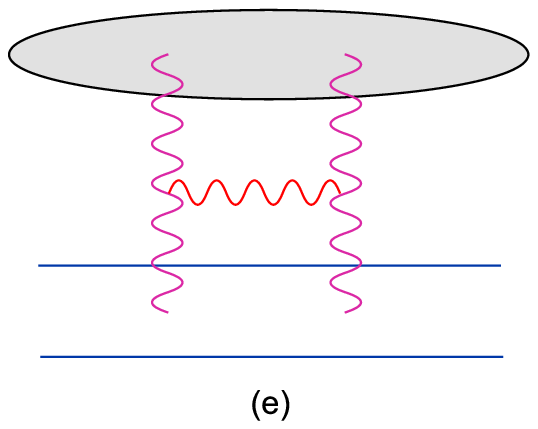}
    \hspace{0.5cm}
    \epsfxsize=5cm\epsfbox{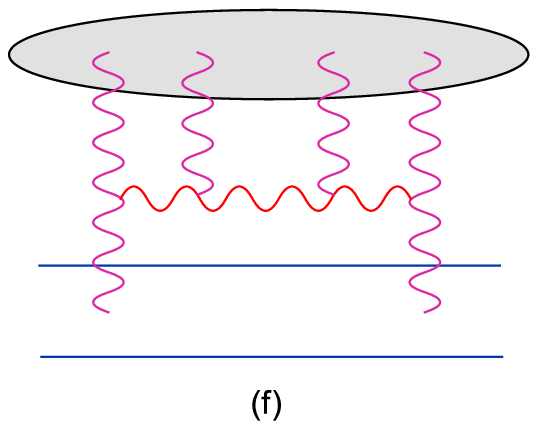}}
    \caption{\sl  Diagrams for single  dipole
    scattering: (a) the tree--level contribution;
    (b,c,d) one step evolution of the projectile;
    (e,f) one step evolution of the target.
    \label{FIG_1DIP}}
    \end{figure}

In Fig. \ref{FIG_1DIP}, we display a few Feynman diagrams which
illustrate the processes encompassed by Eq.~(\ref{Balitsky}) from
the perspective of projectile evolution. These diagrams are proper
to the dipole picture, so the evolution is shown as dipole splitting
rather than as gluon emission. For more clarity, we show only one
diagram contributing to each type of process, which is moreover
taken at the lowest non--trivial order in perturbation theory. Thus,
the scattering between a dipole and the CGC target starts at two
gluon exchange, as shown in Fig. \ref{FIG_1DIP}.a. The one--step
evolution of this amplitude then involves the diagrams in Figs.
\ref{FIG_1DIP}.b, c, d : Fig. \ref{FIG_1DIP}.b is the virtual term
(the original dipole interacts before splitting), Fig.
\ref{FIG_1DIP}.c shows the scattering of one of the daughter dipoles
(there is a similar diagram for the other dipole), while Fig.
\ref{FIG_1DIP}.d describes their simultaneous scattering. We shall
return to a discussion of these diagrams after also introducing the
picture of the target evolution.

\begin{figure}
\begin{center}
\centerline{\epsfig{file=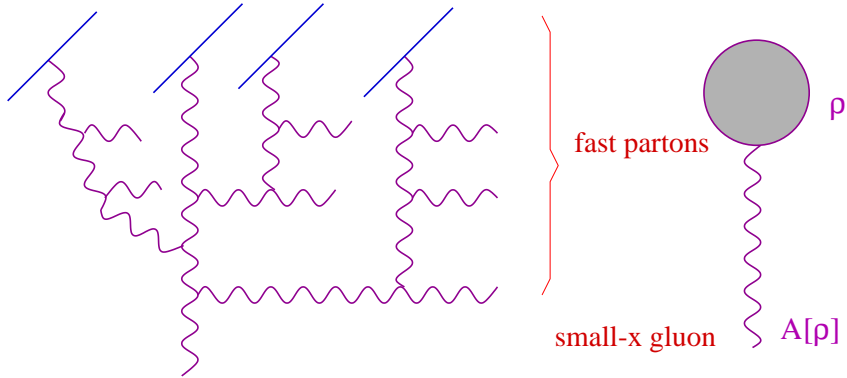,height=5.cm}}
 \caption{\sl A typical gluon cascade which contributes to a
 classical color field configuration in the color glass condensate
\label{CASC}}
\end{center}
\end{figure}

{\bf b)} A different derivation of the Balitsky equations is
provided \cite{RGE} by the color glass condensate (CGC) formalism,
in which the evolution is achieved by boosting the target, and the
non--linear terms correspond explicitly to saturation effects in the
target wavefunction.  The `color glass' is a random superposition of
classical color fields with a gauge--invariant weight function which
evolves with $Y$ according to JIMWLK equation \cite{JKLW97,RGE,W}.
Diagramatically, each classical field configuration corresponds to a
set of {\it gluon cascades} which are initiated by `color sources'
(e.g., valence quarks) carrying a sizeable fraction of the target
rapidity $Y$, and which end up with a small--$x$ gluon at the
comparatively low rapidity of the projectile. (See Fig. \ref{CASC}
for a pictorial representation.) When increasing $Y$, these cascades
evolve through gluon radiation from the classical field created in
the previous steps. For weak fields ($A\simle 1$), corresponding to
low gluon density, the JIMWLK equation reduces to the standard BFKL
evolution: the various $N$--point functions evolve independently
from each other, and rise rapidly with $Y$. But at high density, the
classical fields are strong ($A\gg 1$, and even $A\sim 1/g$ in the
`condensate' regime at saturation), and the dynamics is fully
non--linear.

For the subsequent discussion, it is important to notice that the
non--linearities in the CGC correspond to gluon {\it merging}, but
not also to gluon {\it splitting} : As illustrated in  Fig.
\ref{CASC}, different gluon cascades can fuse with each other, via
the non--linear effects encoded in the classical field equations
(the Yang--Mills equations), and also in the emission and the
propagation of the quantum gluons. This recombination process is
what we refer to as `gluon merging'. On the other hand, the quantum
gluons radiated from the classical fields are not allowed to further
radiate small--$x$ gluons by themselves; that is, the `gluon
splitting' is not included in the evolution. This amounts to
neglecting gluon number fluctuations in the CGC picture. (Indeed,
gluon splitting would be the CGC analog of the dipole splitting
discussed in Sect. \ref{General}.) This is a good approximation at
high density, where gluon splitting is indeed suppressed as compared
to direct emission from the strong classical fields. But as we shall
argue later, gluon splitting is a dominant effect in the dilute
regime at high $k_\perp$, and plays also an essential role in the
approach towards saturation.

This discussion has an obvious consequence for the structure of
non--linearities in the JIMWLK evolution equations for Green's
functions: gluon merging can only {\it reduce} the number of active
fields in a correlation function. Therefore, the r.h.s. of the
equation for the $N$--point function involves in general all the
$n$--point functions with $n\ge N$, but not also those with $n< N$
(compare in this respect with Eq.~(\ref{dn20}), where the opposite
situation happens).

Consider now the scattering between the CGC and an external
projectile which is a collection of dipoles. The scattering
amplitude for a single dipole can be computed in the eikonal
approximation as
 \be\label{Tdipole}
T({\bm{x}},{\bm{y}})\,=\,1\,-\,\frac{1}{N_c}\,
 {\rm tr}\big(V^\dagger({\bm{x}}) V({\bm{y}})\big)
 ,\ee
where $V^\dagger({\bm{x}})$ and $V({\bm{y}})$ are Wilson lines
describing the scattering of the quark and, respectively, antiquark
component of the dipole, e.g. (with P denoting path--ordering in
$x^-$) :
\begin{equation}
\label{Vdef} V^\dagger({\bm{x}}) \,\equiv\,{\rm P}\,{\rm
exp}\left({\rm i}g\int dx^- A^+_a(x^-,{\bm{x}}) t^a\right),
\end{equation}
and $A^+_a$ is the longitudinal component of the color field in the
target (the only dynamical field left in the problem), whose
correlations are described by the CGC. The evolution equation for
$\langle T(1)T(2)\cdots T(N)\rangle_Y$ is then obtained by using the
definition (\ref{Tdipole})--(\ref{Vdef}) of $T$ together with the
equations satisfied by the field correlators $\langle A^+(1)
A^+(2)\cdots\rangle_Y$,  which in turn follow from the JIMWLK
equation\footnote{In practice, it is often simpler to work out
directly the evolution equations for the correlations of the Wilson
lines. But thinking in terms of field correlations helps clarifying
the diagrammatic interpretation of the JIMWLK evolution.}. As
anticipated, the ensuing equations are precisely those originally
obtained by Balitsky, and which at large $N_c$ are generated by
Eq.~(\ref{Balitsky}).

Let us now describe the diagrammatic interpretation of these
equations from the perspective of target evolution. Once again, we
show only a minimal set of representative diagrams. To easy read
these diagrams, notice that the lowest--order contribution to the
scattering amplitude (\ref{Tdipole}), as obtained after expanding
the Wilson lines there, is the 2--gluon exchange:
 \be\label{Tdipole2}
T({\bm{x}},{\bm{y}})\,\approx\,\frac{g^2}{4N_c}\,\big(A^+_a({\bm{x}})
-A^+_a({\bm{y}})\big)^2 + {\mathcal O}(g^3)\,.
 \ee
The two diagrams shown in Figs. \ref{FIG_1DIP}.e and f correspond to
the linear and non--linear terms in Eq.~(\ref{Balitsky}),
respectively. The diagram in Fig. \ref{FIG_1DIP}.e represents the
first step in the BFKL evolution of the 2--point function $\langle
A^+ A^+\rangle_Y$ (the `unintegrated gluon distribution'); in terms
of projectile evolution, it corresponds to the diagrams shown in
Figs. \ref{FIG_1DIP}.b and c. But Fig. \ref{FIG_1DIP}.f exhibits a
non--linear effect which goes beyond BFKL evolution: four gluons
merge into two (a 4--point function reduces to a 2--point function)
through a vertex which at large--$N_c$ can be
recognized\footnote{Indeed, according to Eq.~(\ref{Tdipole2}), the
four gluons which enter this vertex from the above are pairwise
coupled into two color singlets, so like the two gluons emerging
from the vertex; after subsequent evolution, any of these 2--gluon
singlet exchanges would be eventually converted into BFKL ladders,
or `pomerons'.} as the `triple pomeron vertex'. Fig.
\ref{FIG_1DIP}.f corresponds to Fig. \ref{FIG_1DIP}.d for projectile
evolution; both diagrams involve the same triple pomeron vertex,
only with a different interpretation: {\it i)} In Fig.
\ref{FIG_1DIP}.d, this vertex describes the {\it splitting} of one
dipole into two dipoles which then scatter {\it both} with the
target (so that the net effect of the evolution is to replace a
single--pomeron exchange, cf. Fig. \ref{FIG_1DIP}.a, by a
double--pomeron exchange, Fig. \ref{FIG_1DIP}.d). {\it ii)} In Fig.
\ref{FIG_1DIP}.f,  the same vertex describes gluon {\it merging} in
the target wavefunction.

\begin{figure}[t]
   \centerline{\epsfxsize=5cm\epsfbox{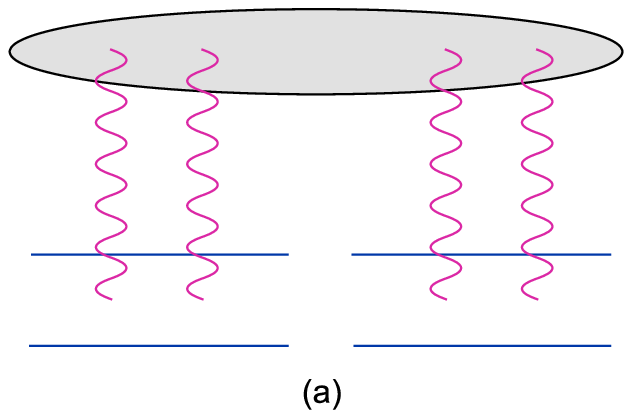}
    \hspace{0.5cm}
    \epsfxsize=5cm\epsfbox{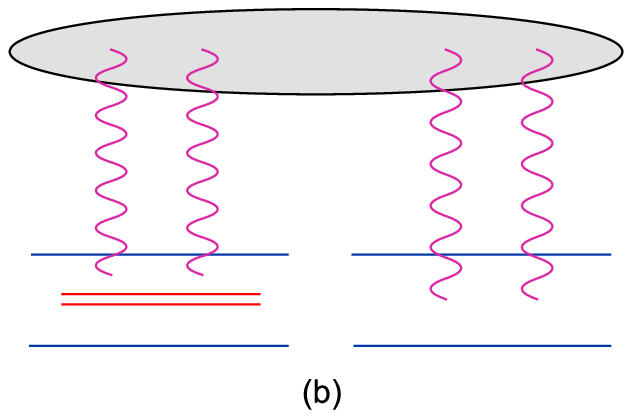}}
    \vspace{1cm}
    \centerline{\epsfxsize=5cm\epsfbox{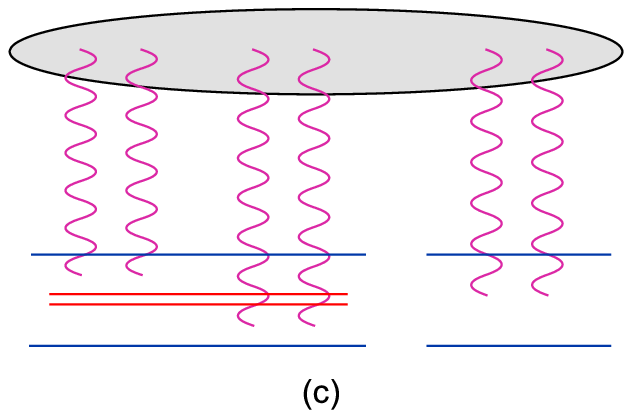}
    \hspace{0.5cm}
    \epsfxsize=5cm\epsfbox{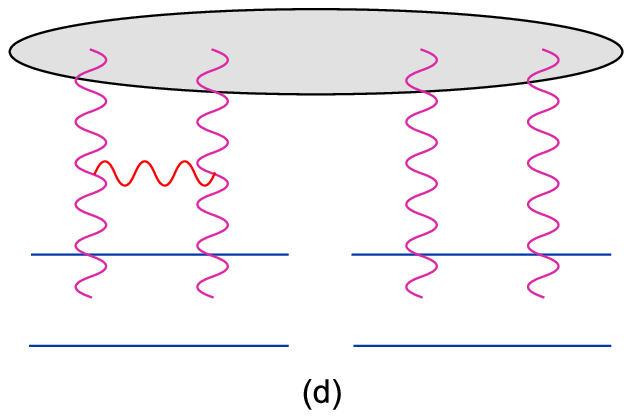}}
    \vspace{1cm}
    \centerline{\epsfxsize=5cm\epsfbox{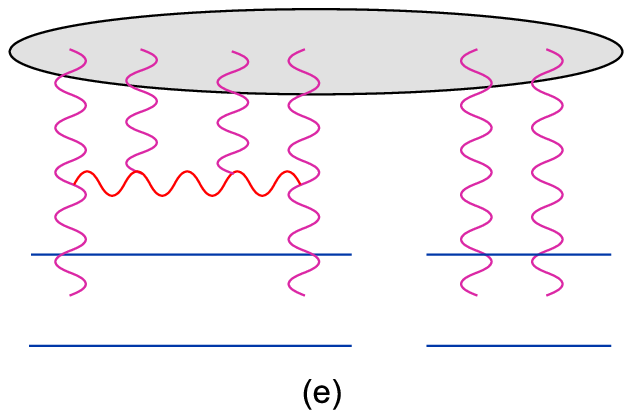}
    \hspace{0.5cm}
    \epsfxsize=5cm\epsfbox{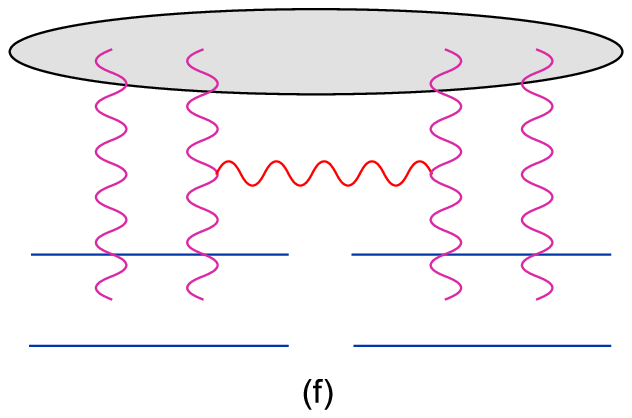}}
    \vspace{1cm}
    \centerline{\epsfxsize=5cm\epsfbox{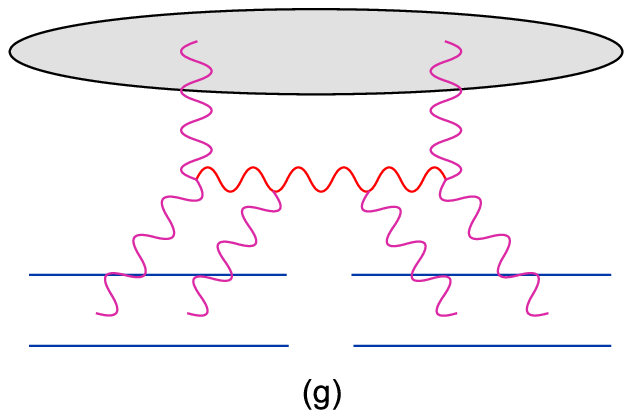}}
    \caption{\sl Diagrams for dipole pair scattering:
    (a) tree--level contribution;
    (b, c) one step projectile evolution;
    (d, e) one step target evolution;
    (f) a diagram suppressed at large $N_c$;
    (g) the missing diagram, which is of leading order
    in both $\alpha_s$ and $N_c$.
    \label{FIG_2DIP}}
\end{figure}

By  `iterating' the diagrams in Fig. \ref{FIG_1DIP}, it is
straightforward to deduce the diagrammatic interpretation of the
equation satisfied by $\langle T^{(N)}\rangle_Y$ for any $N\ge 1$.
Although this seems like a trivial exercise, we nevertheless exhibit
the corresponding results for $\langle T^{(2)}\rangle_Y$ in Figs.
\ref{FIG_2DIP}.b and c (projectile evolution) and \ref{FIG_2DIP}.d
and e (target evolution), only to emphasize that these are not {\it
all} the diagrams which would be expected within perturbative QCD to
the order of interest. (Fig. \ref{FIG_2DIP}.a shows the
corresponding tree--level diagram.) This lack of perturbative
completeness is to be related to our previous remarks concerning the
lack of saturation effects in the projectile evolution and,
respectively, the lack of gluon number fluctuations in the target
evolution. As we shall shortly see, both omissions refer in fact to
the same physical process, only seen from different points of view.

Specifically, in Figs. \ref{FIG_2DIP}.f and g we show two diagrams
which are formally of the same order in $\alpha_s$ as the diagrams
for {\it linear} evolution in Figs. \ref{FIG_2DIP}.b and d, yet they
were not included in the previous analysis. For the diagram in Fig.
\ref{FIG_2DIP}.f, this absence is easy to explain: this diagram is
subleading at
large $N_c$, 
so it is naturally missed by Eq.~(\ref{Balitsky}); as a matter of
facts, this diagram {\it is} correctly included  in the full version
of Balitsky equations (valid for arbitrary $N_c$), and also in the
JIMWLK formalism. As for the second diagram, Fig. \ref{FIG_2DIP}.g,
this is easily recognized to describe the incriminated process,
which is missing in either Balitsky, or JMWLK, picture of the
non--linear evolution: Seen from the perspective of projectile
evolution, this diagram describes the interaction between two
dipoles in the projectile, that is, a {\it saturation effect},
whereas from the perspective of the target, it rather corresponds to
{\it gluon splitting}.

By inspection of the diagrams in Fig. 4, one can understand what was
the argument for ignoring such contributions in the JIMWLK evolution
of the target, and also when is this argument expected to break
down. Consider the dependence of the processes in Fig.
\ref{FIG_2DIP} upon the gluon density in the target: A 2--point
function counts like the density, a 4--point function, like the
density squared, etc. Then the standard, BFKL, diagram in Fig.
\ref{FIG_2DIP}.d is proportional to the gluon density squared, the
recombination diagram in Fig. \ref{FIG_2DIP}.e scales like the third
power of the density times an additional factor $\alpha_s$, while
the splitting diagram, Fig. \ref{FIG_2DIP}.g, scales likes
$\alpha_s$ times the density\footnote{Overall, there is the same
power of $\alpha_s$ associated with the diagrams in  Figs.
\ref{FIG_2DIP}.d and \ref{FIG_2DIP}.g. But when interpreting these
diagrams in terms of gluon densities, one should remember that one
factor of $\alpha_s$ is included in the definition of the gluon
density.}. Clearly, this last contribution is indeed negligible
whenever the density is large enough. For instance, in the
saturation region, the density is ${\mathcal O}(1/\alpha_s)$, so
both the linear diagram, Fig. \ref{FIG_2DIP}.d, and the
recombination diagram, Fig. \ref{FIG_2DIP}.e, contribute on equal
footing, whereas the splitting diagram, Fig. \ref{FIG_2DIP}.g, is
totally irrelevant. However, the situation changes in the dilute
regime at high transverse momenta, where the density is very low
(the average gluon occupation number can be much smaller than one);
then, all diagrams are suppressed {\it except} for the splitting
one,  Fig. \ref{FIG_2DIP}.g, which shows how a 4--point correlation
gets generated via the splitting of a 2--point one. Clearly, this is
the dominant mechanism for constructing higher--point correlation
functions in the dilute regime.

To summarize this discussion, the Balitsky--JIMWLK equations include a
single type of triple pomeron vertex, which corresponds to either
pomeron splitting, or pomeron merging, depending upon the
perspective that we have over the evolution. Still, both merging
{\it and} splitting are necessary in order to have a complete
description of the approach towards saturation, which is consistent
with perturbative QCD. Among the approaches to high energy evolution
proposed so far, the dipole picture includes only splitting (and
thus can be used to describe particle number fluctuations at
high--$k_\perp$), whereas the JIMWLK evolution contains only merging
(so it describes correctly the saturation effects at high density).
As explained in the Introduction, our goal is to combine the
respective virtues of these two approaches in order to achieve
a more complete description of the high--energy evolution of the
scattering amplitudes. Our strategy can be summarized as follows:

Since the physical regimes in which the interesting phenomena ---
saturation and fluctuations (or merging and splitting)
--- play important roles are kinematically well separated, it should be
possible to use a `piecewise' description in which each regime is
covered by a different formalism. Such a description should
correctly capture the essential physics that we are interested in,
although it would probably fail to reproduce the fine details of the
transition between the two regimes. Specifically, in what follows we
shall stick to {\it target evolution}, and we shall rely on the
JIMWLK formalism in the high--density regime, and on the dipole
picture in the dilute one (which means, in particular, that we shall
limit ourselves to the large--$N_c$ approximation). Thus, both gluon
merging, Fig. \ref{FIG_2DIP}.e, and gluon splitting, Fig.
\ref{FIG_2DIP}.g, will be encoded in the target wavefunction,
although their respective theoretical descriptions will be
different. But, as we shall see, these descriptions can be naturally
merged with each other in the equations for the scattering
amplitudes, thus leading to a generalization of the Balitsky--JIMWLK
hierarchy which includes the effects of gluon number fluctuations in
the dilute regime, and thus generates pomeron loops\footnote{The
simplest `pomeron loop' is a one--loop diagram generated by a (BFKL)
pomeron which splits into two which then recombine with each other.
Thus, for this diagram to exists, one needs the simultaneous
presence of triple pomeron vertices for splitting and merging.}
through iterations.

\section{A toy model borrowed from statistical physics}
\label{Toy} \setcounter{equation}{0}


A main difficulty in our present analysis is the lack of a unified
theoretical framework for describing both saturation and particle
number fluctuations in the hadron lightcone wavefunction. It is therefore
instructive, before we embark ourselves in a more detailed study of
QCD, to make a detour through a simple particle model borrowed from
statistical physics which contains both physical ingredients
(particle number fluctuations and recombination) in a unified
setting. This model is formulated similarly to the QCD dipole
picture that we shall discuss in the next section --- there is a
system of particles whose distribution evolves in time according to
a master equation --- but it goes beyond the dipole picture by
including recombination effects (so, in that respect, it may be
viewed as a non--linear version of the dipole picture).

For this model, we shall study the effects of fluctuations on the
evolution equations for particle number correlations, and, in particular,
we shall show that it is possible to  {\it exactly} reformulate the
dynamics as a specific Langevin equation.
This is interesting since the Langevin equation is better suited for
numerical studies than the original master equation.
Inspired by this, we shall later be able to also reformulate
the equations for scattering amplitudes in QCD as a corresponding
Langevin equation. Besides, some of the conceptual and technical
issues that we shall clarify in this section will appear again later,
in the context of QCD.

For more clarity, in this section we shall consider the simplest
version of the statistical model: a zero--dimensional system (no
spatial dimension) in which the only interesting distribution is
that of the particle number. In the Appendix, we shall then discuss
a more refined model (taken from Refs. \cite{PL99,PS01}),
which has one spatial dimension, and which in spite of its
relative simplicity leads to a Langevin equation very
similar to the one that we shall find in QCD. This is the
sFKPP equation \cite{Saar,Panja} (and Refs. therein).

Specifically, in this section we consider a system whose state is
described by the number $n$ of particles A. One can think of all these
particles as being located at some unique lattice site, in which
case $n$ represents the occupation number of that site. The dynamics
is determined by two random elementary processes: under an
infinitesimal step $dY$ in time, a particle can {\it split} into two
with a rate $\alpha$ :
\begin{equation}
        {\rm A} \xrightarrow[]{\alpha} {\rm A} + {\rm A},
\end{equation}
while two particles can {\it recombine} into one with a rate $2\beta$
:
\begin{equation}
    {\rm A} + {\rm A} \xrightarrow[]{2\beta} {\rm A},
\end{equation}
Clearly, the probability $P(n)=P(n,Y)$ to find a configuration with
$n$ particles changes with time, and the corresponding {\it master
equation} reads
\begin{equation}\label{toymaster}
    \frac{d P(n)}{dY} \,=\,
    \frac{d P(n)}{dY} \bigg|_{\alpha} +  \frac{d P(n)}{dY}
    \bigg|_{\beta}\,,
\end{equation}
where the first term in the r.h.s. is the contribution of splitting:
\begin{equation}\label{toyprob}
    \frac{d P(n)}{dY} \bigg|_{\alpha}=
    \alpha \left[(n-1) P(n-1) - n P(n)\right],
\end{equation}
while the second one is due to recombination:
\begin{equation}\label{toyprobrec}
    \frac{d P(n)}{dY} \bigg|_{\beta}=
    \beta \left[n (n+1) P(n+1) - n (n-1) P(n)\right].
\end{equation}

 The expectation value of an observable $\mathcal{O}(n)$ is
given by
\begin{equation}
    \lan \mathcal{O}(n) \ran_Y =
    \sum_n P(n,Y)\,\mathcal{O}(n),
\end{equation}
and its evolution equation can be obtained in general from
Eq.~(\ref{toymaster})--(\ref{toyprobrec}).
For example, it is not hard to find how the
average values of the ``number operators'' $\lan n \ran$, $\lan n^2
\ran$,... change with time. We have
 \be \frac{d \lan n \ran}{dY}&=&
    \alpha \lan n \ran - \beta \left[\lan n^2 \ran -\lan n \ran\right], \nn
    \frac{d \lan n^2 \ran}{dY}&=&
    \alpha \left[ 2 \lan n^2 \ran + \lan n \ran \right]
    - \beta \left[\lan n^3 \ran -3\lan n^2 \ran + \lan n
    \ran\right], ...
\ee We shall also need the {\it normal--ordered} number operators,
defined as
\begin{equation}\label{order}
    :\!n^\kappa\!:= n (n-1) \cdot\cdot\cdot (n-\kappa +1).
\end{equation}
To understand why we call this a ``normal--ordered product", imagine
a second--quantized formalism in which particles of type $A$ are
created, or annihilated, by standard Fock space operators,
$a^\dagger$ and, respectively, $a$, with $[a^\dagger,\,a]=1$. Then
the particle number operator is $n=a^\dagger a$, and we have $n^2
=a^\dagger aa^\dagger a= a^\dagger a^\dagger aa + a^\dagger a =
:\!n^2\!: + n$, so $:\!n^2\!: = n(n-1)$, so like in
Eq.~(\ref{order}). Note furthermore that $:\!n^2\!:$ gives zero when
acting on a state with a single particle (unlike $n^2$, which yields
one). Thus, in the present context, the normal--ordering in
$:\!n^2\!:$ is introduced in order for this operator to properly
count the pairs of particles (so that, e.g., a single particle
cannot be counted like a pair), and similarly for $:\!n^\kappa\!:$ .
Moreover, as we shall see in the next section, and also in the
Appendix, the use of normal ordering is the only way to avoid
equal--point singularities for systems with spatial dimensions.

 The evolution equation for $n^{(\kappa)} \equiv
\langle :\!n^{\kappa}\!: \rangle$ takes a rather simple form:
\be\label{toyevoln}
    \frac{d n^{(\kappa)}}{dY}&=&
    \alpha \left[
    \kappa \,n^{(\kappa)} + \kappa (\kappa-1)\, n^{(\kappa-1)}
    \right] -\beta \left[
    \kappa \,n^{(\kappa+1)} + \kappa (\kappa-1)\, n^{(\kappa)}
    \right]\,.
 \ee
But the simplicity is rather illusory, as $n^{(\kappa)}$ is coupled in
the evolution with $n^{(\kappa-1)}$ and $n^{(\kappa+1)}$, so
Eq.~(\ref{toyevoln}) is just a particular equation within an infinite
hierarchy. In what follows, we shall study various aspects of this
hierarchy: In particular, we shall discuss the respective roles of
the multiplication and recombination processes, and also the
relative importance of fluctuations versus mean field aspects at
different stages of the evolution. But before doing that, let us
first introduce a different stochastic process which is
mathematically equivalent to the one that we have studied so far, in
the sense of generating exactly the same correlations as obtained by
solving the hierarchy in Eq.~(\ref{toyevoln}).

Specifically, let us consider the following Langevin equation
\begin{equation}\label{toylang}
    \frac{d\tilde{n}}{ dY} =  \alpha\, \tilde{n}
    -\beta\, \tilde{n}^2
    +\sqrt{2 \left[ \alpha\, \tilde{n}
    -\beta\, \tilde{n}^2 \right]}\, \nu
    \equiv A\, +B\nu\,,
\end{equation}
where $\nu(Y)$ is a Gaussian white noise : $\lan \nu(Y)\ran=0$ and
$\lan \nu(Y)\nu(Y')\ran = \delta(Y-Y'),$ and the right hand side
must be understood with the Ito prescription for  time
discretization (recall that a Langevin process is not
differentiable): Namely, if one writes $Y=j\epsilon$, where $j=0,1,2,\dots$
and $\epsilon$ is the length of the time step, then
Eq.~(\ref{toylang}) should be properly understood as
\begin{equation}\label{toylangdiscr}
    \frac{\tilde{n}_{j+1}-\tilde{n}_j}{\epsilon} =  \alpha\, \tilde{n}_j
    -\beta\, \tilde{n}_j^2
    +\sqrt{2 \left[ \alpha\, \tilde{n}_j
    -\beta\, \tilde{n}^2_j \right]}\, \nu_{j+1}
    \equiv A_j\, +B_j\,\nu_{j+1},
\end{equation}
with
 \be \lan\nu_j\ran = 0,\qquad  \lan \nu_i \nu_j\ran =
\frac{1}{\epsilon}\,\delta_{ij}\,,\ee
 which shows that updating the variables $\tilde{n}$ at
time--step $j+1$ only requires knowledge of the variables at
time--step $j$. The noise term in Eq.~(\ref{toylang}) is said to be
{\it multiplicative} since it is multiplied by a function of
$\tilde{n}$. It is easy to see that $\tilde{n}=0$
is an {\it unstable} fixed point for the evolution described by
Eq.~(\ref{toylang}), whereas $\tilde{n}= \alpha/\beta$ is a {\it
stable} fixed point, and corresponds to saturation. If the initial
condition satisfies $0\le \tilde{n}(0)\le \alpha/\beta$, this will
remain true for all subsequent times.

For an arbitrary function $F(\tilde{n})$, Eq.~(\ref{toylang})
implies the following evolution equation
\begin{equation}\label{toyF}
    \frac{d \lan F(\tilde{n}) \ran}{dY}=
    \lan A F'(\tilde{n}) \ran
    +\frac{1}{2} \lan B^2 F''(\tilde{n}) \ran.
\end{equation}
By taking $F(\tilde{n})=\tilde{n}^\kappa$, it is straightforward to
show that the hierarchy emerging from the Langevin problem is indeed
equivalent to the hierarchy of the toy model, Eq.~(\ref{toyevoln}),
provided we identify $\lan \tilde{n}^\kappa \ran$ with
$n^{(\kappa)}$. In particular, the noise term in
Eq.~(\ref{toylangdiscr}) is responsible for the {\it fluctuation}
terms in Eq.~(\ref{toyevoln}), i.e., the terms proportional to
$\kappa (\kappa-1)$ within the square brackets.

It is in fact easy to check directly at the level of the original
hierarchy, Eq.~(\ref{toyevoln}), that $n^{(\kappa)} =
(\alpha/\beta)^\kappa$ is indeed a fixed point of the evolution.
Moreover, in the presence of fluctuations, we do not expect other
fixed points (see also below). Thus, for sufficiently large $Y$, the
occupation number will saturate to $n_{\rm sat} = \alpha/\beta$
independently of the initial conditions. We would like this
saturation value to be large, $n_{\rm sat} \gg 1$, in order for the
analogy with QCD to be as close as possible. We shall therefore
choose $ \alpha\gg \beta$, which implies that the recombination
effects (proportional to $\beta$) will be parametrically suppressed
as compared to the growth effects (proportional to $\alpha$) except
in the saturation regime --- a situation similar to QCD. Thus, if
one starts in a dilute regime ($n\ll n_{\rm sat}$) at $Y=0$, then it
is possible to follow the evolution in the early stages (prior to
saturation) by using the {\it linearized} version of
Eq.~(\ref{toyevoln}), as obtained after neglecting recombination :
\be\label{toylinevoln}
    \frac{d n^{(1)}}{dY}\,=\,
    \alpha \,n^{(1)},\qquad
    \frac{d n^{(2)}}{dY}\,=\,
    2\alpha \left[
    n^{(2)} + \, n^{(1)}
    \right],\quad\dots
\ee As we shall see in Sect. \ref{Dipoles}, the above equations are
the analog of the equations satisfied by the dipole densities in the
dipole picture. Note that these equations are still coupled with
each other, because of the {\it fluctuation} terms: The term linear
in $n^{(1)}$ in the r.h.s. of the evolution equation for $n^{(2)}$
describes a fluctuation in which one particle splits into two (so
this is the analog of the dipole splitting term in
Eq.~(\ref{dn20})). One may expect fluctuations not to be important
at large times where $ n^{(2)} \gg  n^{(1)}$, but in general this is
not right. To see this, consider the solution to
Eqs.~(\ref{toylinevoln}) with generic initial conditions
$n^{(1)}(0)=n_0$ and $n^{(2)}(0)=n^{(2)}_0$ :
\begin{align}\label{toysol1}
    n^{(1)}(Y) &= n_0 \exp(\alpha Y),
    \nn 
    n^{(2)}(Y) &=
    \left[
    n^{(2)}_0 + 2 n_0
    \right] \exp(2 \alpha Y)
    -2 n_0 \exp(\alpha Y),\,\,\dots
\end{align}
An interesting initial condition is $n_0=1$ and $n^{(\kappa)}_0 =0$
for $\kappa\ge 2$, meaning that there is a single particle in the
initial state. (In QCD, this would correspond to the case where one
starts the evolution with a single dipole, or, more generally, with
an isolated fluctuation at high $k_\perp$.) Then
Eqs.~(\ref{toysol1}) show that so long as $\alpha Y \lesssim 1$,
$n^{(1)}$ dominates over $n^{(2)}$ in the r.h.s. of the second
equation (\ref{toylinevoln}), and thus is driving force for the
growth in the number of pairs. {\it In the absence of fluctuations,
the pair number $n^{(2)}$ would remain zero for ever.} For larger
$Y$, $n^{(2)}$ starts to dominate over $n^{(1)}$ (as it increases
faster), but the ratio
\begin{equation}\label{toyratio}
    \frac{n^{(2)}(Y)}{[n^{(1)}(Y)]^2} = 2 - 2 \exp(-\alpha Y),
\end{equation}
approaches 2, and thus differs from the naive prediction
$n^{(2)}(Y)\approx [n^{(1)}(Y)]^2$ of the mean field approximation.
In fact, one can check on Eqs.~(\ref{toysol1}) that the only
way to satisfy this mean--field factorization at intermediate values
of $Y$ (prior to saturation) is to assume a large occupation number
already in the initial conditions: $n_0\gg 1$, $n^{(2)}_0 \approx
n_0^2 \gg n_0$, etc. (In QCD, this would correspond, e.g., to
starting the evolution with a very large nucleus at $Y=0$.)

This conclusion is not altered by the addition of the recombination
terms, which merely provide saturation: For the model at hand and
for generic initial conditions ($n_0\sim{\mathcal O}(1)$), the mean
field approximation is justified only in the saturation regime
achieved at very large $Y$ : $Y\simge Y_0\simeq
(1/\alpha)\ln(\alpha/\beta)$. For a more general system which
involves also spatial dimensions, so like QCD or the
reaction--diffusion model to be studied in the Appendix, particles
can escape from the bulk, so there will be always a region in
phase--space in which the density is low and the mean field
approximation is not applicable, however large is $Y$ and
independently of the initial conditions (even if one starts with a
large nucleus): this is the tail of the distribution, in which the
evolution is driven by fluctuations.

In view of the comparison with QCD, it is useful to notice that the
analog of the (dipolar) Balitsky--JIMWLK equations in the present toy model
are the equations obtained after neglecting the fluctuation terms in
Eq.~(\ref{toyevoln}), that is
 \be\label{toyBal}
    \frac{d n^{(\kappa)}}{dY}\,=\,
    \kappa \left[
     \alpha\,n^{(\kappa)}  \,-\,\beta\,n^{(\kappa+1)}
     \right]\qquad{\rm (no\,\,fluctuation)}\,.\ee
(Alternatively, the analog of  Eq.~(\ref{Balitsky}) --- the
generating equation of the Balitsky--JIMWLK hierarchy --- is the
deterministic equation obtained after ignoring the noise term in
the Langevin equation (\ref{toylang}).)
Neglecting fluctuations in the {\it recombination} term is indeed
harmless, since this term is important only in the high--density
regime, where fluctuations become irrelevant anyway. But for the
{\it growth} term, which acts also in the dilute regime, the
fluctuations {\it are} crucial, as shown by the previous analysis.
In particular, the reduced hierarchy in Eq.~(\ref{toyBal}) could never
generate a high density system from an initial state which involves
only one particle.

Another effect of the fluctuations is to wash out some remarkable
properties of Balitsky equations, like the existence of exact
factorized solutions \cite{JP04,Janik} :
Eq.~(\ref{toyBal}) can be solved with the
Ansatz $n^{(\kappa)}=c_{\kappa}\,[n^{(1)}]^\kappa$ provided the
coefficients $c_{\kappa}$ satisfy $c_{\kappa}=c_2^{\kappa-1}$, in
agreement with Ref. \cite{Janik}. (In particular, $n^{(\kappa)} =
(1/c_2)(\alpha/\beta)^\kappa$ is a one--parameter family of fixed
points for  Balitsky equations.) But it can be easily checked that
such a factorized solution does not exist for the complete
equations including fluctuations, that is, Eq.~(\ref{toyevoln}).

More generally, we expect the fluctuations to wash out any
sensitivity to the initial conditions after a sufficiently large
evolution. For the zero--dimensional model in this section, this is
simply the statement that at large $Y$ the solution will converge to
an unique fixed point, namely $n^{(\kappa)} =
(\alpha/\beta)^\kappa$, independently of the initial state. For more
complicated systems, which involve also spatial dimensions, we
expect {\it universality} at late times (in the sense of
insensitivity to the initial conditions) also for more complex
aspects of the dynamics, like the behavior of high energy scattering
amplitudes in QCD.

\section{Fluctuations and evolution in the dipole picture}
\label{Dipoles} \setcounter{equation}{0}

We now return to QCD with a discussion of particle number
fluctuations in the dipole picture. To that aim, we shall not rely
on the original formulation of the dipole picture due to Mueller
\cite{AM94,AM95}, but rather on its alternative formulation in Ref.
\cite{IM032} (see also \cite{LL03}), which is better suited for a
study of fluctuations. The main difference between these two
formulations is that, loosely speaking, `they put the evolution at
different ends' (see also Figs. \ref{Fig_T1} and \ref{Fig_T1_st}
below). More precisely, in
Mueller's original formulation, the evolution proceeds via splitting
at the highest rapidity end: the rapidity increment $dY$ is used to
accelerate the {\it original} dipole (the one which has initiated
the evolution), which then undergoes an additional splitting, whose
effects propagate in the whole configuration. But in this picture
the correlation between the high--rapidity dipole which has split
and the low--rapidity one that we measure is distributed over many
steps of rapidity, and thus is difficult to trace back. By contrast,
in the formulation in Ref. \cite{IM032}, the splitting occurs in one
of the dipoles at the lowest rapidity end, which was itself
generated in the previous step of the evolution, and which is the
direct parent of the dipole we measure. This makes it easier to
follow correlations associated with splitting, which are
particularly important in the first few steps after the splitting
occurs.

Following Ref. \cite{IM032}, the system of dipoles generated by the
evolution up to rapidity $Y$ of an original dipole with coordinates
$\bm{x}_0$ and $\bm{y}_0$ will be described as a stochastic ensemble
of dipole configurations endowed with a probability law which
evolves with $Y$ according to a {\it master equation}. Specifically,
a given configuration is specified by the number of dipoles $N$ and
by  $N-1$ transverse coordinates $\{\bm{z}_i\}=\{\bm{z}_1,
\bm{z}_2,...\bm{z}_{N-1}\}$, such that the coordinates of the $N$
dipoles are $(\bm{z}_0,\bm{z}_1)$,
$(\bm{z}_1,\bm{z}_2)$,...,$(\bm{z}_{N-1},\bm{z}_N)$, with $\bm{z}_0
\equiv \bm{x}_0$ and $\bm{z}_N \equiv \bm{y}_0$. The probability
$P_N(\{\bm{z}_i\};Y)$ to find a given configuration obeys the
following evolution equation (this is similar to Eq.~(\ref{toyprob})) :
\begin{align}\label{evolP}
    \frac{\del P_N(\bm{z}_1,...\bm{z}_{N-1};Y)}{\del Y}
    =&-\frac{\abar}{2\pi}
    \left[
    \sum_{i=1}^N \int \!d^2 \bm{z}\,
    \mathcal{M}(\bm{z}_{i-1},\bm{z}_i,\bm{z})
    \right]
    P_N(\bm{z}_1,...\bm{z}_{N-1};Y)
    \nonumber \\
    &+\frac{\abar}{2\pi}
    \sum_{i=1}^{N-1}
    \mathcal{M}(\bm{z}_{i-1},\bm{z}_{i+1},\bm{z}_i)
    \,P_{N-1}
    (\bm{z}_1,...,\bm{z}_{i-1},\bm{z}_{i+1},...,\bm{z}_{N-1};Y),
\end{align}
where we have defined the shorthand notation for the dipole kernel
\begin{equation}\label{dipkernel}
    \mathcal{M}(\bm{x},\bm{y},\bm{z})=
    \frac{(\bm{x}-\bm{y})^2}
    {(\bm{x}-\bm{z})^2 (\bm{y}-\bm{z})^2}\,.
\end{equation}
The expectation value of an operator $\mathcal{O}$ which depends
only the dipole sizes is given by
\begin{equation}\label{opeave}
    \lan \mathcal{O}(Y) \ran \,=\, \sum_{N=1}^\infty\int \!d {\it
    \Gamma}_N\,P_N(\{\bm{z}_i\};Y)\, \mathcal{O}_N(\{\bm{z}_i\}),
\end{equation}
where the phase space integration is simply $d{\it
\Gamma}_N\,=\,{\rm d}^2\bm{z}_1{\rm d}^2\bm{z}_2\dots{\rm
d}^2\bm{z}_{N-1}$. Then by using the master equation (\ref{evolP})
one can show that
\begin{align}\label{DODY}
    \frac{\del \lan \mathcal{O}(Y) \ran}{\del Y}=& \frac{\abar}{2\pi}
    \sum_{N=1}^\infty \int d {\it \Gamma}_N \,
    P_N(\{\bm{z}_i\};Y)\,
    \sum_{i=1}^N
    \int \! d^2\bm{z}\,
    {\mathcal{M}}({\bm{z}}_{i-1},{\bm{z}}_i,{\bm z})
    \nonumber \\
    & \qquad\times\big[- {\cal O}_N(\{\bm{z}_i\})
    + {\cal O}_{N+1}(\{\bm{z}_i,\bm{z}\}) \,\big],
\end{align}
where the $\bm{z}$ argument in ${\cal O}_{N+1}$ is to be placed
between $\bm{z}_{i-1}$ and $\bm{z}_i$.

In what follows, we shall use Eq.~(\ref{DODY}) to derive evolution
equations for the dipole number densities. Because of the difference
alluded to before with respect to the original formulation of the
dipole picture, the final equations that we shall obtain are not the
same as the corresponding equations in the early literature on the
dipole picture (see Refs. \cite{AM94,AM95,Salam95,AMSalam96}),
although the two sets of equations are mathematically equivalent (in
the sense of providing identical results for identical initial
conditions). Very recently, Levin and Lublinsky \cite{LL04} have
obtained the same equations that we shall derive here, but they used
a more straightforward approach, based on a generating functional
\cite{LL03}, which however hides some of the subtle points that we
shall emphasize below, and which in our opinion are important to
properly understand the results.

Consider first the average dipole number density. The corresponding
operator for a $N$-dipole configuration is
\begin{equation}\label{densityN}
    n_{N}(\bm{x},\bm{y})=
    \sum_{j=1}^N
    \delta^{(2)}(\bm{z}_{j-1}-\bm{x})
    \delta^{(2)}(\bm{z}_j-\bm{y}).
\end{equation}
With ${\cal O}_{N}\equiv n_N$, the expression within the square
bracket in the second line of Eq.~(\ref{DODY}) becomes
\begin{align}\label{Deltai}
    \Delta_i(\bm{x},\bm{y},\bm{z}) =
    &-\delta^{(2)}(\bm{z}_{i-1} -\bm{x})\,
    \delta^{(2)}(\bm{z}_{i} -\bm{y})
    \nonumber \\
    &+\delta^{(2)}(\bm{z}_{i-1} -\bm{x})\,
    \delta^{(2)}(\bm{z} - \bm{y})
    +\delta^{(2)}(\bm{z} - \bm{x})\,
    \delta^{(2)}(\bm{z}_{i} -\bm{y}).
\end{align}
After simple manipulations we arrive at the following evolution
equation for the average of the dipole number density
$n_Y(\bm{x},\bm{y})\equiv \lan n(\bm{x},\bm{y})\ran_Y$ :
\begin{align}\label{evolnumber}
    \frac{\del n_Y(\bm{x},\bm{y})}{\del Y}=
    \frac{\abar}{2\pi} \int \! d^2\bm{z}\,
    \big[
    -&\, {\cal M} ({\bm{x}},{\bm{y}},{\bm{z}})\,n_Y(\bm{x},\bm{y})
    \nonumber \\
    +&\, {\cal M} ({\bm{x}},{\bm{z}},{\bm{y}})\,n_Y(\bm{x},\bm{z})
    + {\cal M} ({\bm{z}},{\bm{y}},{\bm{x}})\,n_Y(\bm{z},\bm{y})
    \big]
    \nonumber\\
    &\hspace{-2.8cm}
    \equiv \frac{\abar}{2\pi}\,
    \int \! d^2\bm{z}\,
    {\cal K}_{\bm{x}\bm{y}\bm{z}} \otimes n_Y(\bm{x},\bm{y}).
\end{align}
As anticipated, this is not the same as the standard equation for
the dipole number density in the dipole picture. In the latter, the
splitting occurs in the original dipole $(\bm{x}_0, \bm{y}_0)$, so
the more complete notation $n_Y(\bm{x},\bm{y}|\bm{x}_0,\bm{y}_0)$ is
needed for the dipole density. Then the standard equation reads
 \be\label{nYBFKL}\hspace*{-.7cm}
 \frac{\del n_Y(\bm{x},\bm{y}|\bm{x}_0,\bm{y}_0)}
{\del Y}&=&\frac{\abar}{2\pi} \int \! d^2\bm{z}\, {\cal
M}({\bm{x}_0},{\bm{y}_0},{\bm{z}})\big[-
n_Y(\bm{x},\bm{y}|\bm{x}_0,\bm{y}_0)\nn &{}&\qquad \qquad\qquad
+\,\, n_Y(\bm{x},\bm{y}|\bm{x}_0,\bm{z}) +\,
n_Y(\bm{x},\bm{y}|\bm{z},\bm{y}_0)\big],\ee and is also recognized
as the dipole version of the BFKL equation \cite{BFKL}. The two
equations (\ref{evolnumber}) and (\ref{nYBFKL}) are illustrated in
Figs. \ref{Fig_T1} and \ref{Fig_T1_st},
respectively. In spite of the formal differences,
these equations are nevertheless equivalent\footnote{This can be
checked by using the symmetry property $(\bm{x}-\bm{y})^4
\,n_Y(\bm{x},\bm{y}|\bm{x}_0,\bm{y}_0)\,=\,(\bm{x}_0-\bm{y}_0)^4 \,
n_Y(\bm{x}_0,\bm{y}_0|\bm{x},\bm{y})$.}, as shown in Ref.
\cite{IM032}.

\begin{figure}[t]
    \centerline{\epsfxsize=13cm\epsfbox{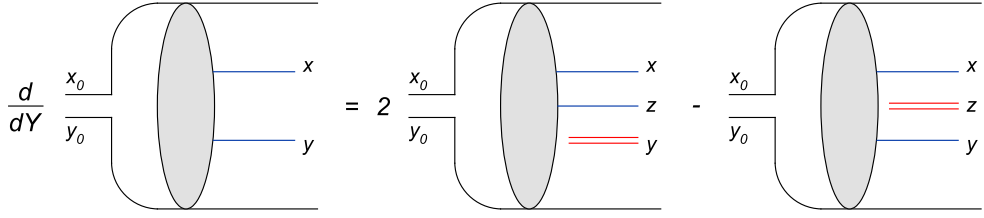}}
    \caption{\sl The one--step evolution of the average dipole
number density as described by Eq.~(\ref{evolnumber}).
 \label{Fig_T1}}
\end{figure}

\begin{figure}[t]
\vspace{0.5cm}
    \centerline{\epsfxsize=13cm\epsfbox{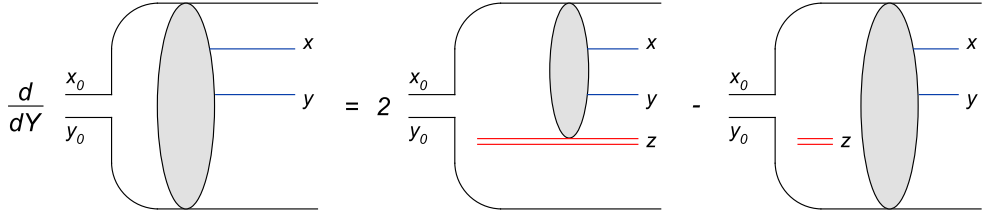}}
    \caption{\sl The same as in Fig. \ref{Fig_T1}, but from the
 perspective of Eq.~(\ref{nYBFKL}).
 \label{Fig_T1_st}}
\end{figure}

Now we turn to the dipole pair density. As in the toy model
considered in the previous section, the corresponding operator for a
given $N$--dipole configuration is defined as the {\it
normal--ordered} product of single densities:
\begin{align}\label{n2def}
    n_N^{(2)}(\bm{x}_1,\bm{y}_1 ; \bm{x}_2,\bm{y}_2)
    = &\! :\! n_N(\bm{x}_1,\bm{y}_1)\,
    n_N(\bm{x}_2,\bm{y}_2) \!:
    \\
    \equiv &\,n_N(\bm{x}_1,\bm{y}_1)
    n_N(\bm{x}_2,\bm{y}_2)
    \!-\!\delta^{(2)}(\bm{x}_1-\bm{x}_2)
    \delta^{(2)}(\bm{y}_1-\bm{y}_2)
    n_N(\bm{x}_1,\bm{y}_1). \nonumber
\end{align}
In fact, it is precisely this normal--ordered operator that is
automatically reproduced by the generating functional used by
Mueller \cite{AM94,AM95}, and also by Levin and Lublinsky
\cite{LL04,LL03}, but the importance of normal ordering has not been
emphasized before, nor the difficulties arising when using the
simple
product (as opposed to the normal--ordered one) 
have been discussed.

The use of normal ordering is required by both physical and
technical considerations\footnote{We are grateful to Larry
McLerran for his insightful observations on the role of normal
ordering in this particular context.}.
On the physical side, note that
Eq.~(\ref{n2def}) is equivalent to a restricted double sum:
\begin{align}\label{n2def1}
    n_N^{(2)}(\bm{x}_1,\bm{y}_1 ; \bm{x}_2,\bm{y}_2)=\!
    \sum_{\substack{j,k=1\\j \neq k}}^N
    \delta^{(2)}(\bm{z}_{j-1}-\bm{x}_1)
    \delta^{(2)}(\bm{z}_j-\bm{y}_1)
    \delta^{(2)}(\bm{z}_{k-1}-\bm{x}_2)
    \delta^{(2)}(\bm{z}_k-\bm{y}_2),
\end{align}
which does not allow for ``pairs'' made of the same dipole. On the
technical side, note that the subtraction performed in the second
line of Eq.~(\ref{n2def}) eliminates the `ultraviolet' singularity
of the simple product $n_N(\bm{x}_1,\bm{y}_1)n_N(\bm{x}_2,\bm{y}_2)$
in the limit where the two dipoles have identical
coordinates\footnote{This is manifest on the expression in
Eq.~(\ref{n2def1}) which is well defined when $\bm{x}_1=\bm{x}_2$
and $\bm{y}_1=\bm{y}_2$. The only dangerous term in this limit would
have been the term with $j=k$ in the unrestricted double sum.}.
Because of that, the evolution equation for the normal--ordered pair
density is well defined in the continuum limit, unlike for the
simple product.

 Specifically, by using Eq.~(\ref{n2def}), we see that the square
bracket in the second line of Eq.~(\ref{DODY}) becomes (recall also
Eq.~(\ref{Deltai}))
\begin{align}\label{deltan2}
  n_{N+1}^{(2)} - n_N^{(2)}\!= &\,
    \Delta_i(\bm{x}_1,\bm{y}_1,\bm{z})
    \,n_N(\bm{x}_2,\bm{y}_2)
    +\Delta_i(\bm{x}_2,\bm{y}_2,\bm{z})
    n_N(\bm{x}_1,\bm{y}_1)
     \\
    +\,&\Delta_i(\bm{x}_1,\bm{y}_1,\bm{z})\,
    \Delta_i(\bm{x}_2,\bm{y}_2,\bm{z})
    -\delta^{(2)}(\bm{x}_1-\bm{x}_2)
    \,\delta^{(2)}(\bm{y}_1-\bm{y}_2)\,
    \Delta_i(\bm{x}_1,\bm{y}_1,\bm{z}).\nonumber
\end{align}
The contribution of the first two terms to the evolution equation is
rather easy to obtain and reads (to simplify writing, we use
the notation introduced in Eq.~(\ref{evolnumber}) together with
$n^{(2)}_Y\equiv \lan n^{(2)}\ran_Y$)
\begin{equation}
    \frac{\abar}{2\pi}\,
    \int \! d^2\bm{z}\,
    \left[
    {\cal K}_{\bm{x}_1\bm{y}_1\bm{z}}+
    {\cal K}_{\bm{x}_2\bm{y}_2\bm{z}}
    \right]
    \otimes
    n^{(2)}_Y(\bm{x}_1,\bm{y}_1;\bm{x}_2,\bm{y}_2)
    \, +{\,\rm linear},
\end{equation}
where the linear terms not explicitly shown arise from
normal--ordering the product $ n_N(\bm{x}_1,\bm{y}_1)\,
n_N(\bm{x}_2,\bm{y}_2)$. By themselves, these linear terms are
divergent when $\bm{x}_1=\bm{x}_2$ and/or $\bm{y}_1=\bm{y}_2$, but
in the complete calculation they exactly cancel the singular terms
coming from $\Delta_i(\bm{x}_1,\bm{y}_1,\bm{z})\,
\Delta_i(\bm{x}_2,\bm{y}_2,\bm{z})$ and from the last term in
Eq.~(\ref{deltan2}). All the remaining terms are non--singular, and
add up to give the following contribution to $n_{N+1}^{(2)} -
n_N^{(2)}$ :
 \begin{equation}
    \left[
    \delta^{(2)}(\bm{z}_{i-1}-\bm{x}_1)\,
    \delta^{(2)}(\bm{z}-\bm{y}_1)\,
    \delta^{(2)}(\bm{z}-\bm{x}_2)\,
    \delta^{(2)}(\bm{z}_{i}-\bm{y}_2)
    \right]
    \,+\, \big\{1 \leftrightarrow 2\big\},
\end{equation}
which in turn gives the following contribution, linear in the dipole
number density $n_Y$, to the evolution equation for the dipole pair
density :
\begin{equation}
    \frac{\abar}{2\pi}\,
    \mathcal{M}({\bm{x}}_1,\bm{y}_2,\bm{x}_2)\,
    n_Y(\bm{x}_1,\bm{y}_2)\,
    \delta^{(2)}(\bm{x}_2-\bm{y}_1)
    \,+\,\big\{1 \leftrightarrow 2\big\}.
\end{equation}
Putting everything together we finally arrive at
\begin{align}\label{eqn2}
    \frac{\del\, n_Y^{(2)}(\bm{x}_1,\bm{y}_1 ; \bm{x}_2,\bm{y}_2)}
    {\del Y} =
    \frac{\abar}{2\pi}\,
    \bigg[
    &\int \! d^2 \bm{z} \,
    {\cal K}_{\bm{x}_1\bm{y}_1\bm{z}}
    \otimes
    n^{(2)}_Y(\bm{x}_1,\bm{y}_1;\bm{x}_2,\bm{y}_2)
    \\
    +\,& \mathcal{M}({\bm{x}}_1,\bm{y}_2,\bm{x}_2)\,
    n_Y(\bm{x}_1,\bm{y}_2)\,
    \delta^{(2)}(\bm{x}_2-\bm{y}_1)
    \bigg]
    \,+\, \big\{1 \leftrightarrow 2\big\}.\nonumber
\end{align}
The terms proportional to $n^{(2)}_Y$ in the r.h.s. of the above
equation describe the normal BFKL evolution of the pair density,
while the terms linear in $n_Y$ are recognized as fluctuations in
which the two measured dipoles arise from the splitting of the same
parent dipole (and thus are contiguous with each other). These are
precisely the fluctuations discussed in Sect. \ref{General}. As
explained there, and also for the statistical model in
Sect. \ref{Toy}, the main effect of these fluctuations is to drive
the growth of pair density in the dilute regime where $n_Y$
dominates over $n^{(2)}_Y$ in the r.h.s. of Eq.~(\ref{eqn2}).

\begin{figure}[t]
    \centerline{\epsfxsize=13cm\epsfbox{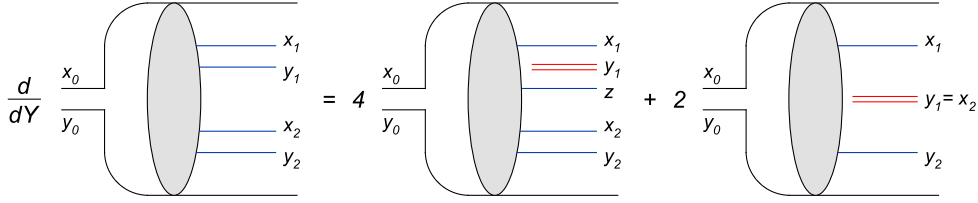}}
    \caption{\sl The one--step evolution of the dipole pair density
$n_Y^{(2)}$ as described by Eq.~(\ref{eqn2}). The virtual BFKL term
is not shown. \label{Fig_T2}}
\end{figure}

The evolution described by Eq.~(\ref{eqn2}) is illustrated in Fig.
\ref{Fig_T2}. Once again, this equation is not the same as the standard
equation for $n^{(2)}_Y$ which appears in the literature
\cite{AM94,AM95,Salam95,AMSalam96} (but it coincides with the
corresponding equation of Levin and Lublinsky \cite{LL04}). The
standard equation is rather obtained by putting the evolution in the
original dipole (so like in Eq.~(\ref{nYBFKL})), and its r.h.s.
involves $n^{(2)}_Y$ together with terms {\it quadratic} in $n_Y$,
but not also terms linear in $n_Y$. Fluctuations are therefore more
difficult to recognize on that equation.

For later use it is also convenient to express the fluctuation term
in terms of the dipole sizes $\bm{r}_i ={\bm{x}}_i-\bm{y}_i$ and the
corresponding impact parameters $\bm{b}_i=({\bm{x}}_i+\bm{y}_i)/2$,
instead of the coordinates of their quark and anti--quark legs. One
thus finds
\begin{equation}\label{fluctrb}
    \frac{\abar}{2\pi}\,
    \frac{(\bm{r}_1+\bm{r}_2)^2}
    {\bm{r}_1^2\, \bm{r}_2^2}\,
    n_Y \left(
    \bm{r}_1+\bm{r}_2,
    \bm{b}_1-\frac{\bm{r}_2}{2}
    \right)\,
    \delta^{(2)} \left(
    \bm{b}_1-\bm{b}_2-\frac{\bm{r}_1+\bm{r}_2}{2}
    \right)
    \,+\, \big\{1 \leftrightarrow 2\big\},
\end{equation}
which is the same as Eq.~(\ref{dn20}), as expected.

Note finally that via the same manipulations as above it is
straightforward to construct the equation obeyed by the $k$--body
dipole density $n^{(k)}_Y$ within the dipole picture. This is the
analog of the $\alpha$--piece of Eq.~(\ref{toyevoln}) in the
statistical model, that is, it involves a piece proportional to
$n^{(k)}_Y$ which represents the normal BFKL evolution, and a piece
linear in  $n^{(k-1)}_Y$ which describes fluctuations and dominates
at low density. The corresponding equations can be found in Ref.
\cite{LL04}.

\section{Fluctuations and saturation in dipole--CGC
scattering} \label{Langevin} \setcounter{equation}{0}

In the previous section, we have seen that the dipole picture provides
a natural theoretical framework to describe gluon number fluctuations
in the dilute regime, and for large $N_c$. In what follows, we shall
exploit the relation between dipole densities and scattering
amplitudes in order to transfer the effects of fluctuations from the
target wavefunction to the evolution equations for the amplitudes. In
the dilute regime, where the fluctuations are important, the
scattering amplitudes are simply proportional to the dipole densities
in the target (see Sect. 6.1), and it is therefore straightforward to
translate the previous equations in Sect. \ref{Dipoles} into
corresponding equations for the amplitudes. The ensuing equations can
be then extended to the high--density regime by
simply adding the same non--linear terms as in the Balitsky--JIMWLK
equations (since the fluctuations cannot change significantly the
non--linear aspects of the dynamics). These manipulations will lead
us, in Sect. 6.2, to a new set of equations which include both
fluctuations and saturation, and which are the main result in this
paper.  Finally, in Sect. 6.3, we shall show that,  mediating a
coarse--graining in impact parameter space, this infinite set of
equations can be replaced by a single Langevin equation, which looks
formally like the Kovchegov equation supplemented by a noise term.

\subsection{From dipole densities to scattering amplitudes}

Let us start with the scattering between a single external dipole
(``the projectile") and a dilute target whose
wavefunction is described in the dipole picture. The very
statement that ``the target is dilute" does not refer to the target
{\it alone}, but rather is a statement about how the target looks
like on the resolution scale of the projectile, which is its
transverse size. Namely, this implies that the size $r$ of the
external dipole is small as compared to the local saturation length
$1/Q_s(Y,\bm{b})$ in the target at the dipole impact parameter $\bm{b}$. 
Under these conditions, the scattering with the external dipole  
acts as a measure of
the density of gluons (or dipoles) with transverse size $r$ in the
target wavefunction. This is so since the dipole--dipole scattering
is {\it quasi--local in transverse phase--space}, as we recall now:

A priori, the external dipole can scatter off any of the dipoles
present in the target wavefunction, so its scattering amplitude can
be evaluated as : 
 \be\label{Tonscatt}
T(\bm{r},\bm{b},Y) =\int \! d^2 \bm{r}_1 \,\int \! d^2 \bm{b}_1
\,{\mathcal T}_0 (\bm{r},\bm{r}_1,\bm{b}-\bm{b}_1)\, n
 (\bm{r}_1,\bm{b}_1,Y),\ee
where $ n (\bm{r}_1,\bm{b}_1,Y)$ is the number density of dipoles
with size $\bm{r}_1$ and impact parameter $\bm{b}_1$ in the target
wavefunction, as produced after a rapidity evolution $Y$, and
${\mathcal T}_0 (\bm{r},\bm{r}_1,\bm{b}-\bm{b}_1)$ is the amplitude
for the scattering between two elementary dipoles.
Eq.~(\ref{Tonscatt}) holds for a given configuration of the target
wavefunction, that is, for a given scattering event. The
corresponding expression for the {\it average} amplitude is obtained
after averaging over all the configurations in the target, as
explained in Sect. \ref{Dipoles}.

Note that in writing Eq.~(\ref{Tonscatt}) we have restricted
ourselves to a single--scattering approximation, as appropriate in
the dilute regime. This equation is correct only so long as
$T(\bm{r},\bm{b},Y)\ll 1$. For consistency, it is sufficient to
evaluate ${\mathcal T}_0$ to lowest order in perturbation theory,
that is, to ${\mathcal O} (\alpha_s^2)$, which corresponds to the
exchange of two gluons. The corresponding expression is well known
in the literature, but here we shall need only its general
properties. Namely, ${\mathcal T}_0\sim \alpha_s^2$ when the two
dipoles have similar sizes and are relatively close to each other in
impact parameter space, but it decreases very fast, as ${\mathcal
T}_0 (\bm{r},\bm{r}_1,\bm{b}-\bm{b}_1) \sim \alpha_s^2 r^2 r_1^2/ (
\bm{b}-\bm{b}_1)^4$, for a large separation $|\bm{b}-\bm{b}_1|\gg
r_>$ between the two dipoles (with $r_>={\rm max} (r,r_1)$). One can
thus replace $\bm{b}_1 \to \bm{b}$ as the argument of the density
$n$ in the r.h.s. of Eq.~(\ref{Tonscatt}), and at the same time
evaluate the integral over $\bm{b}_1$ as:
 \be
 \int \! d^2
 \bm{b}_1 \,{\mathcal T}_0 (\bm{r},\bm{r}_1,\bm{b}-\bm{b}_1)\,\simeq\,
 \int_{r_>} \! d^2 \bm{b}_r \,\frac{\alpha_s^2 r^2 r_1^2}{
 \bm{b}_r^4}\, \simeq \,\alpha_s^2 r^2_<\,,\ee
where $\bm{b}_r=\bm{b}-\bm{b}_1$ and $r_<={\rm min} (r,r_1)$.

To evaluate also the remaining integral over $\bm{r}_1$, one needs
to know the corresponding dependence of the dipole density. So long
as we are in the genuine BFKL regime, characterized by a
non--trivial ``anomalous dimension", the dominant behavior of
$n(\bm{r}_1,\bm{b},Y)$ is expected to be of the form
 \be\label{nBFKL}
 n(\bm{r}_1,\bm{b},Y)\,\sim\,\frac{1}{\bm{r}_1^4}\,\big(
 \bm{r}_1^2 Q_c^2(\bm{b},Y)\big)^{\gamma_0}\,,
 \ee
where $\gamma_0\approx 0.63$, $1-\gamma_0$ is the anomalous dimension,
and $Q_c^2(\bm{b},Y)\propto {\rm e}^{\lambda \bar\alpha_s Y}$ is a
line of constant density (up to the kinematical factor $1/
\bm{r}_1^4$) in the $\ln(1/r_1)-Y$ plane. (This line is parallel to
the saturation line.) Strictly speaking, Eq.~(\ref{nBFKL}) holds for
the {\it average} density, but for sufficiently large $Y$, we expect
the dipole distribution to self--average even in a single event, and
therefore show the same gross features as the average distribution.
After inserting Eq.~(\ref{nBFKL}) in the r.h.s. of
Eq.~(\ref{Tonscatt}), one can easily check that the ensuing integral
is saturated by $r_1\sim r$, so, by dimensional arguments, the final
result reads simply:
 \be\label{Tn1}
 T(\bm{r},\bm{b},Y) \,\simeq \,\alpha_s^2 r^4
 n(\bm{r},\bm{b},Y)\,.\ee
In the dilute BFKL regime in which we are interested here, this
equation is exact up to some numerical fudge factor of order one,
that we do not control in the above approximation, and which should
depend upon the value of the anomalous dimension.

Eq.~(\ref{Tn1}) can be immediately extended to the scattering of two
or more external dipoles. In a particular event, the scattering
amplitude for a system of several external dipoles is simply the
product of the individual amplitudes for each dipole. For instance,
 \comment{ \be\label{Tonscatt2}
\Tt_Y(\bm{r}_1,\bm{b}_1;\bm{r}_2,\bm{b}_2) &=& \frac{1}{2} \int \!
d^2 \bm{r}_a d^2 \bm{b}_a \,d^2 \bm{r}_b d^2 \bm{b}_b \,
n_Y(\bm{r}_a,\bm{b}_a) \,n_Y(\bm{r}_b,\bm{b}_b) \nn &{}&\qquad
\times\,{\mathcal T}_0 (\bm{r}_1,\bm{r}_a,\bm{b}_1-\bm{b}_a)\,
 \,{\mathcal T}_0 (\bm{r}_2,\bm{r}_b,\bm{b}_2-\bm{b}_b)\nn &=&
 T(\bm{r}_1,\bm{b}_1,Y)\, T(\bm{r}_2,\bm{b}_2,Y)\,
 ,\ee
where the factor $1/2$ in the r.h.s. takes care of the symmetry of
the integral under the exchange of the two dipoles $a$ and $b$ from
the target, and using similar manipulations as before, one finds}
 \be\label{Tn2}\hspace*{-.5cm}
 \Tt_Y(\bm{r}_1,\bm{b}_1;\bm{r}_2,\bm{b}_2)  &=&
 T(\bm{r}_1,\bm{b}_1,Y)\,T(\bm{r}_2,\bm{b}_2,Y)
 \nn &\simeq& \alpha_s^4\,
 r^4_1 r^4_2\,:\!n(\bm{r}_1,\bm{b}_1,Y) n(\bm{r}_2,\bm{b}_2,Y)\!:
\,,\ee
where the final expression is correct up to some unknown fudge factor.
Note that this final expression involves the {\it normal--ordered}
dipole pair density, as defined in
Eqs.~(\ref{n2def})--(\ref{n2def1}). This is so since we neglect here
the possibility that both external dipoles scatter off the same
dipole in the target. As explained in Sect. \ref{General}, this is a
good approximation so long as the external dipoles have different
sizes and/or impact parameters.

\subsection{The equations for the scattering amplitudes}
\label{NewEqs}

 By using Eqs.~(\ref{Tn1}) and (\ref{Tn2}) together
with the equations for the average one--body and two--body dipole
densities in the target, as established in Sect. \ref{Dipoles}, one
can immediately deduce the equations satisfied by the average
scattering amplitudes within the present approximations.

For the scattering amplitude of a single dipole $\langle T({\bm
 r},{\bm b}) \rangle_Y\equiv \langle T({\bm x},{\bm y}) \rangle_Y$,
we thus find the BFKL equation, as expected:
 \be\label{T1BFKL}
 {\partial \langle T({\bm
 x},{\bm y})\rangle_Y\over {\partial Y}} &=& \frac{\bar\alpha_s}{2\pi}\int d^2{\bm z}
 \,{\cal M} ({\bm{x}},{\bm{y}},{\bm{z}})
 \big\langle -T({\bm x},{\bm y})+T({\bm x},{\bm z})+T({\bm
 z},{\bm y})\big\rangle_Y\nn
 &\equiv& \frac{\abar}{2\pi}\,
    \int \! d^2\bm{z}\,
    {\cal M}_{\bm{x}\bm{y}\bm{z}} \otimes \langle T({\bm
  x},{\bm y})\rangle_Y.
 \ee
Note that after multiplication by the fourth power of the dipole
size, the various dipole kernels appearing in the original equation
(\ref{evolnumber}) for the dipole density have been all converted
into the same kernel ${\cal M} ({\bm{x}},{\bm{y}},{\bm{z}})$.

Eq.~(\ref{T1BFKL}) applies so long as $T\ll 1$. But we know already
what should be the corresponding generalization to the high density
regime where $T\sim 1$ : In that regime, the target wavefunction is
described by the CGC formalism, which should not be significantly
altered by fluctuations, because the latter are relatively
unimportant when the density is large. Thus, the general equation
for $\langle T\rangle$ can be simply obtained by adding to
Eq.~(\ref{T1BFKL}) the appropriate non--linear term generated by the
JIMWLK evolution of the target. As discussed in Sect. \ref{BJIMWLK},
this non--linear term is $\lan -T({\bm x},{\bm z})T({\bm z},{\bm
y})\ran_Y$. With this addition, Eq.~(\ref{T1BFKL}) becomes identical
to the first equation (\ref{EQT1}) in the Balitsky hierarchy.

Consider now the scattering amplitude for two dipoles $\lan \Tt
({\bm x}_1,{\bm y}_1;{\bm x}_2,{\bm y}_2)\ran_Y \\ \equiv \lan
T({\bm x}_1,{\bm y}_1)\, T({\bm x}_2,{\bm y}_2)\ran_Y$. In the
dilute regime, this is obtained by combining Eqs.~(\ref{Tn2}) and
(\ref{eqn2}), while in the high density regime this should also
include the non--linear terms (involving $\lan T^{(3)}\ran_Y$)
induced by the JIMWLK evolution. By putting all these ingredients
together, we find the following evolution equation (with the concise
notation introduced in the second line of Eq.~(\ref{T1BFKL}))
 \be\label{T2final}\hspace*{-.7cm}
 {\partial \left\langle T^{(2)}({\bm x}_1,{\bm y}_1;{\bm x}_2,{\bm y}_2)
 \right\ran_Y
 \over {\partial Y}} &=&\frac{\bar\alpha_s}{2\pi}\Big\{\int d^2{\bm z}
 \,\Big({\cal M}_{\bm{x}_1\bm{y}_1\bm{z}} \otimes
 \left\langle T^{(2)}({\bm x}_1,{\bm y}_1;{\bm x}_2,{\bm y}_2)\right\ran_Y
 \nn&{}&\qquad -\,{\cal M} ({\bm{x}_1},{\bm{y}}_1,{\bm{z}})\,
 \left\lan T^{(3)}({\bm x}_1,{\bm z}; {\bm z},{\bm y}_1;{\bm x}_2,{\bm y}_2)
 \right\rangle_Y\Big)
 \nn&{}&\qquad+\,\kappa\,\alpha_s^2\,
    \frac{\bm{r}_1^2\, \bm{r}_2^2}{(\bm{r}_1+\bm{r}_2)^2}\,
    \big \lan T ({\bm x}_1,{\bm y}_2)\big\ran_Y\,
    \delta^{(2)} \left({\bm x}_2-{\bm y}_1
    \right)\Big\}\nn &{}& 
    \,+\, \big\{1 \leftrightarrow 2\big\}
 .
 \ee
Among the various terms in the r.h.s. of Eq.~(\ref{T2final}), those
in the first two lines were already present in the corresponding
Balitsky equation,
while the term linear in $\lan T \ran_Y$ in the third line (that we
have written in the mixed notations of Eq.~(\ref{dn20})) is a new
term, which takes into account the effect of fluctuations.
The unknown fudge factor $\kappa\sim {\cal O}(1)$ in this term has
been introduced to parametrize our uncertainty concerning the
precise relation between $T$ and $n$, cf. Eqs.~(\ref{Tn1}) and
(\ref{Tn2}). Note that this uncertainty does not affect the other,
more standard, terms in the r.h.s. of Eq.~(\ref{T2final}), which
come out the same as in the Balitsky--JIMWLK equations.

Let us discuss the properties of this new term in some detail.
First, this is of order $\alpha_s^2\,\lan T \ran$, so it is
parametrically suppressed when $T\gg \alpha_s^2$ (since in that case
already the uncorrelated piece $\lan T \ran\lan T \ran$ of $\lan
T^{(2)}\ran$ is much larger than the fluctuation term). In
particular, the new term is certainly unimportant in the saturation
regime, where Eq.~(\ref{T2final}) reduces to the corresponding
Balitsky equation. On the other hand, the fluctuation term dominates
in the very dilute regime where $\lan T \ran\ll \alpha_s^2$, and
thus is responsible for the growth of $\lan T^{(2)}\ran$ in that
regime. For instance, if we start with a target made of a single
dipole at $Y=0$, then $\lan T^{(2)}\ran_0=0$, and the rise of $\lan
T^{(2)}\ran$ in the early stages of the evolution is driven by the
fluctuation term. A similar discussion applies to a rare
high--$k_\perp$ (or small size) fluctuation generated in the
wavefunction of an arbitrary target. We conclude that the usual BFKL
evolution is not applicable in the dilute regime, in contrast to
what one could naively expect. This failure will be further analyzed
in Sect. \ref{Physics}.

Consider also the geometry of the fluctuation term in the transverse
plane, as manifest on Eq.~(\ref{T2final}). As anticipated in Sect.
\ref{General}, for the external dipoles to feel the effect of
fluctuations, they must be contiguous with each other, i.e., ${\bm
x}_2={\bm y}_1$ or ${\bm y}_2={\bm x}_1$. Moreover, for contiguous
dipoles of unequal transverse sizes, the importance of the
fluctuation is controlled by the size of the smallest dipole:
${\bm{r}_1^2  \bm{r}_2^2}/{(\bm{r}_1+\bm{r}_2)^2}\sim r_<^2$. This
is so because of the geometry of dipole splitting (see Fig. 1)
together with the fact that, in order to scatter, two dipoles have
to overlap with each other. When a (target) dipole splits into two
dipoles of very different sizes, the small child dipole is
necessarily located near the edge of the parent dipole. Thus, for
the scattering to take place, the impact parameter of the {\it
incoming} small dipole should be located within a distance $r_<$
from the edge of the parent dipole (of size $|\bm{r}_1+\bm{r}_2|\sim
r_>$). This condition introduces the geometrical penalty factor
$r_<^2/r_>^2$ manifest in the fluctuation term in
Eq.~(\ref{T2final}). To conclude, the effects of the fluctuations
are larger for incoming dipoles of comparable sizes.

\begin{figure}[t]
    \centerline{\epsfxsize=6cm\epsfbox{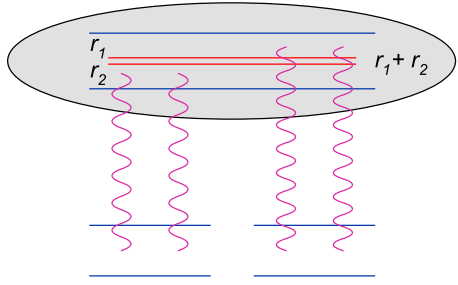}}
    \caption{\sl Diagrammatic illustration of the fluctuation
term in Eq.~(\ref{T2final}) : the original dipole with size
$\bm{r}_1 + \bm{r}_2$ splits
at the time of the interaction into two new dipoles
with  sizes $\bm{r}_1$ and, respectively, $\bm{r}_2$, which then
scatter off two external dipoles.
\label{Fig_split}}\vspace*{.5cm}
\end{figure}

\begin{figure}[t]
    \centerline{\epsfxsize=12cm\epsfbox{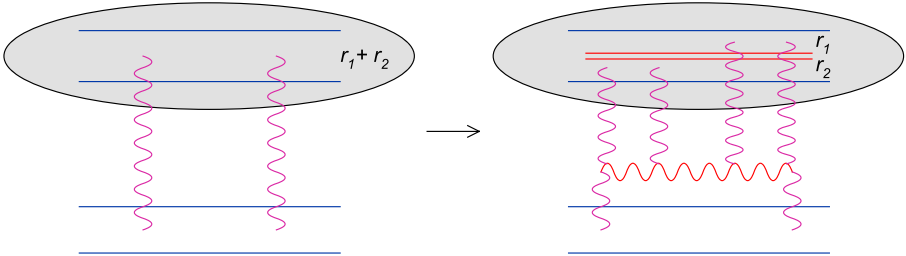}}
    \caption{\sl Two steps in the evolution of the average
scattering amplitude of a single dipole: the original amplitude
(left) and its evolution after two steps (right).
 \label{Two_steps}}
\end{figure}

Consider also the diagrammatic interpretation of the fluctuation
term. This is illustrated in Fig. \ref{Fig_split}, where we
represent the target as a collection of dipoles (as appropriate in
the dilute regime), one out of which splits into two dipoles which
then interact both with the external ones. This diagram is the
dipole picture analog of the 2--to--4 splitting diagram in Fig.
\ref{FIG_2DIP}.g. Of course, the dipole picture cannot be used to
also describe the non--linear effects in the previous equations,
which correspond to saturation effects in the target wavefunction.
To interpret the latter, we should rather resort to the CGC picture.
In Fig. \ref{Two_steps}, we depict a particular two--step evolution
of the single dipole amplitude $\lan T \ran$, as emerging from
Eqs.~(\ref{EQT1}) and (\ref{T2final}). The upper part of this
diagram is the same as the splitting process in Fig. 8. The lower
part describes the recombination of the four gluons resulting from
that splitting into two; this is a diagram of the CGC formalism
(compare to Fig. 2.f) which is encoded in the non--linear term in
Eq.~(\ref{EQT1}). The overall diagram in Fig. 9 represents the
simplest pomeron loop generated by the present equations.

It is straightforward to write down the generalization of
Eq.~(\ref{T2final}) valid for an amplitude $\lan
T^{(k)}\ran$ with $k\ge 3$. This will be the corresponding (dipolar)
Balitsky equation supplemented by fluctuation terms linear in $\lan
T^{(k-1)}\ran$ which account for the possibility that a pair of
dipoles $(\bm{r}_i,\bm{r}_j)$ within $\lan T^{(k)}\ran$ which are
contiguous with each other be generated through the splitting of a
unique dipole of size $\bm{r}_i+\bm{r}_j$ from $\lan
T^{(k-1)}\ran$. The ensuing hierarchy of equations, which
generalizes the Balitsky hierarchy at large $N_c$ by including the
effects of gluon number fluctuations in the dilute regime,
represents our main result in this paper. As already manifest on
Eq.~(\ref{T2final}), the equations in this hierarchy
are afflicted with an uncertainty
concerning the normalization of the fluctuation terms, which
reflects the approximation (\ref{Tn1}) used in their derivation.
As we shall explain in Sect. \ref{Physics}, this uncertainty does not
affect the leading order predictions of these equations at very large
$Y$ and small $\alpha_s$, but it may influence the
subleading effects.

\subsection{The Langevin equation}

The equations for scattering amplitudes constructed in the previous
subsection are similar to, although more complicated than, the
general equations (\ref{toyevoln}) for particle number correlations
in the statistical toy model. In Sect. \ref{Toy}, we have seen that
the hierarchy in Eq.~(\ref{toyevoln}) can be equivalently
represented by a Langevin equation with an appropriate noise term.
In what follows we shall show that, mediating some approximations
which allow one to get rid of the impact parameter dependence of the
amplitudes, a similar Langevin equation can be written also in QCD.

The Langevin equation will be seen to emerge naturally once that we
succeed to rewrite the equations for the scattering amplitudes in a
form {\it quasi--local} in $\bm{b}$. To that aim, we need to assume
that the target is quasi--homogeneous over distances of the order of
the size of the external dipoles. This is a reasonable approximation
so long as we are interested in the local evolution towards
saturation/unitarity, i.e., in the local `blackening' of the target.
In particular, this is sufficient for a study of the energy
dependence of the saturation momentum. On the other hand, this
approximation cannot be used to study the transverse expansion of
the black disk, or to compute total cross--sections.

Under this assumption, all the terms in the previous equations {\it
except for the fluctuation terms} become local in $\bm{b}$. For
instance, the r.h.s. of Eq.~(\ref{EQT1}) for $\langle T({\bm x},{\bm
y}) \rangle\equiv \langle T({\bm r},{\bm b})\rangle$ involves
$T({\bm x},{\bm z})\equiv T({\bm r}',{\bm b} + \frac{{\bm r}-{\bm
r}'}{2})$ (with ${\bm r}'={\bm x}-{\bm z}$), which is approximated
as $T({\bm r}',{\bm b})$. Then, Eq.~(\ref{EQT1}) becomes:
 \be\label{EQT1loc}
&{}&{\partial \langle T({\bm r},{\bm b})\rangle_Y\over {\partial
Y}}\,= \,\frac{\bar\alpha_s}{2\pi}\int d^2{\bm r}'\, {{{\bm
r}^2} \over {{\bm r}'^2 ({\bm r}-{\bm r}')^2}}\\
&{}& \,\,\qquad\quad\big\langle-T({\bm r},{\bm b})+T({\bm r}',{\bm
b})+T({\bm r}-{\bm r}',{\bm b})-T({\bm r}',{\bm b})T({\bm r}-{\bm
r}',{\bm b})\big\rangle_Y,\nonumber
 \ee
which is indeed local in $\bm{b}$. Similarly, all the standard terms
in the equation for  $\lan \Tt ({\bm r}_1,{\bm b}_1;{\bm r}_2,{\bm
b}_2)\ran$ --- that is, the terms in the first two lines in the
r.h.s. of Eq.~(\ref{T2final}) --- become bilocal in $\bm{b}_1$ and
$\bm{b}_2$.
But the fluctuation term there remains non--local, because of the
$\delta$--functions imposing contiguity. We have indeed (compare to
Eq.~(\ref{fluctrb}))
 \be\label{T2fluct}\hspace*{-.7cm}
 {\partial \left\langle T^{(2)}({\bm r}_1,{\bm b}_1;{\bm
 r}_2,{\bm b}_2)\right\ran_Y
 \over {\partial Y}}\Big |_{\rm fluct.}\!
 &=&\,\kappa\, \alpha_s^2\,\frac{\abar}{2\pi}\,
    \frac{\bm{r}_1^2\, \bm{r}_2^2}{(\bm{r}_1+\bm{r}_2)^2}\,
    \big\lan T(\bm{r}_1+\bm{r}_2,
    \bm{b}_1)\big\ran_Y\,\nn &{}&\quad
    \times\, \delta^{(2)} \left(
    \bm{b}_1-\bm{b}_2-\frac{\bm{r}_1+\bm{r}_2}{2}\right)
     \,+\, \big\{1 \leftrightarrow 2\big\}
 ,\ee
where we have approximated $T(\bm{r}_1+\bm{r}_2,
\bm{b}_1-\frac{\bm{r}_2}{2})\approx T(\bm{r}_1+\bm{r}_2, \bm{b}_1)$.

At this level it becomes natural to perform a Fourier transform to
momentum space, by introducing
 \be\label{phiN} \varphi(\bm{k},\bm{b})\,\equiv\,
    \int \frac{d^2 \bm{r}}{2\pi \bm{r}^2}\,
   {\rm e}^{-i \bm{k} \cdot \bm{r}}\, T({\bm r},{\bm b})\,.
    \ee
Up to an overall normalization of ${\cal O}(1/\bar\alpha_s)$,
$\varphi(\bm{k},\bm{b})$ can be
interpreted  as the gluon phase-space occupation number, or the
`unintegrated gluon distribution'  \cite{SCALING,IIT04}. The benefit
of using the momentum space is twofold: (i) After Fourier transform,
the non--linear terms in the evolution equations (e.g., the term
quadratic in $T$ in Eq.~(\ref{EQT1loc})) become {\it local} in
$\bm{k}$. (ii) Since the fluctuation term (\ref{T2fluct}) depends
upon the transverse sizes $\bm{r}_1$ and $\bm{r}_2$ only through
their sum $\bm{r}_1+\bm{r}_2$, its Fourier transform will be {\it
diagonal} in momentum.

Specifically, Eq.~(\ref{EQT1loc}) yields in momentum space  :
 \be\label{EQT1k} \hspace*{-.7cm}
 {\partial \langle \varphi({\bm k},{\bm b})
 \rangle_Y\over {\partial Y}}={\bar\alpha_s}\!\int
 \frac{d^2{\bm p}}{\pi}\, {{{\bm k}^2} \over {{\bm p}^2 ({\bm k}-{\bm p})^2}}
 \Big\langle \,{{\bm p}^2 \over {\bm k}^2}\, \varphi({\bm p},{\bm b}) -
 {1\over 2}\,\varphi({\bm k},{\bm b})\Big\rangle_Y -
 {\bar\alpha_s}\, \big\langle \varphi^2({\bm k},{\bm
 b})\big\rangle_Y,
 \nn
 \ee
where the linear part is recognized as the BFKL equation for the
unintegrated gluon distribution, while the non--linear term provides
saturation.

Furthermore, the Fourier transform of the fluctuating contribution
to $\langle T^{(2)}\rangle$, Eq.~(\ref{T2fluct}), is obtained as
 \be\label{phi2fluct}\hspace*{-.7cm}
 {\partial \left\langle \varphi({\bm k}_1,{\bm b}_1)
 \varphi({\bm k}_2,{\bm b}_2)\right\ran_Y
 \over {\partial Y}}\Big |_{\rm fluct.}\!
 &=&{\abar}\,2\kappa\, \alpha_s^2\,
 \delta^{(2)}(\bm{k}_1-\bm{k}_2)
 \\ &{}&\,\,\times\,\int \frac{d^2 \bm{r}}{2\pi\bm{r}^2}\,
   {\rm e}^{-i \bm{k}_1 \cdot \bm{r}}\,
   \lan T({\bm r},{\bm b}_1)\ran\,
   \delta^{(2)} \left(
    \bm{b}_1-\bm{b}_2-\frac{\bm{r}}{2}\right),\nonumber
 \ee
where the factor of 2 appears because the two terms in the r.h.s. of
Eq.~(\ref{T2fluct}) have given identical contributions. Note that
the exponential in the last integral selects $r\simle 1/k_1$, thus
the fluctuation term is important only for transverse separations $
|\bm{b}_1-\bm{b}_2|\simle 1/k_1$ between the two dipoles.

To also rewrite the fluctuation terms in a form which is local in
$\bm{b}$, we shall proceed to a {\it coarse--graining}. This is
physically motivated, since an external dipole of size $r$ cannot
distinguish details in the target on a transverse scale much smaller
than $r$. That is, the gluon (or dipole) distribution measured by
the external dipole is necessarily averaged in $\bm{b}$ over a disk
of radius $r$ around its impact parameter. After Fourier transform
to momentum space, this corresponds, by the uncertainty principle,
to a coarse--graining over an area $\sim 1/k^2$. To implement this,
we shall divide the impact parameter space into cells of area
$1/k^2$, and average the scattering amplitudes over any such a cell.
Consider a particular cell with center at $\bm{b}_i$. Averaged
quantities within that cell are then defined as:
 \be\label{phii} \varphi_i(\bm{k})\,\equiv\,
   k^2 \int_{\Sigma_i(k)} d^2 \bm{b}\,\varphi(\bm{k},\bm{b}),
   \qquad \varphi_i^2(\bm{k})\,\equiv\,
   k^2 \int_{\Sigma_i(k)} d^2 \bm{b}\,\varphi^2(\bm{k},\bm{b}),\ee
etc., where $\Sigma_i(k)\sim 1/k^2$ is the area of the cell. In
fact, given our previous assumption that the target is
quasi--homogeneous over the area covered by the dipole, it is clear
that the averaging above is tantamount to simply replacing
$\varphi(\bm{k},\bm{b})$ everywhere within a cell by its value at
the center of that cell: $\varphi_i(\bm{k})\approx
\varphi(\bm{k},\bm{b}_i)$. Therefore, $\varphi_i^2(\bm{k})\approx
[\varphi_i(\bm{k})]^2$, so, by itself, this coarse--graining does
not introduce additional correlations.

The coarse--graining is therefore trivial for the standard (BFKL +
non--linear) terms in the evolution equations: it amounts to 
replacing $\varphi({\bm k},{\bm b}) \to \varphi_i(\bm{k})$ (for a
dipole hitting the target in cell $i$) in any of those terms. The
non--locality of the BFKL kernel poses no special difficulties for
this coarse--graining since, e.g., the integral over $\bm{p}$ in
Eq.~(\ref{EQT1k}) is dominated by momenta $p\sim k$. Then
Eq.~(\ref{EQT1k}) implies that the equation for the average
amplitude $\lan\varphi_i(\bm{k})\ran$ in cell $i$ involves the
double--scattering amplitude within the same cell,
$\lan\varphi_i^2(\bm{k})\ran$. To compute this, it is therefore
enough to study the scattering of two external dipoles which hit the
target within the same cell. This is also the interesting case for a
study of fluctuations: as discussed after Eq.~(\ref{phi2fluct}), the
fluctuations vanish when the external dipoles fall in different
cells.

Consider therefore the equation satisfied by $\lan
\varphi_i(\bm{k}_1)\varphi_i(\bm{k}_2)\ran$. The coarse--graining is
non--trivial only for the fluctuation term, in which case it implies
(note that $\bm{k}_1=\bm{k}_2$ in this term, cf.
Eq.~(\ref{phi2fluct})):
   \be &{}& k_1^4\int_{\Sigma_i(k_1)} d^2 \bm{b}_1\,
   \int_{\Sigma_i(k_1)} d^2 \bm{b}_2\,
   \int \frac{d^2 \bm{r}}{2\pi \bm{r}^2}\,
   {\rm e}^{-i \bm{k}_1 \cdot \bm{r}}\,
   \lan T({\bm r},{\bm b}_1)\ran\,
   \delta^{(2)} \left(
    \bm{b}_1-\bm{b}_2-\frac{\bm{r}}{2}\right) \nn
    &{}&\qquad = k_1^4\int_{\Sigma_i(k_1)} d^2 \bm{b}_1\,
    \int \frac{d^2 \bm{r}}{2\pi \bm{r}^2}\,
   {\rm e}^{-i \bm{k}_1 \cdot \bm{r}}\, \lan T({\bm r},{\bm b}_1)\ran\,
    = k_1^2 \lan\varphi_i(\bm{k})\ran.
  \ee
The integration over $\bm{b}_2$ has been used to remove the
$\delta$--function (whose support is indeed included in
$\Sigma_i(k_1)$), while the integration over $\bm{b}_1$ has served
together with Eq.~(\ref{phiN}) to construct
$\lan\varphi_i(\bm{k})\ran$.

Let us summarize here the first two equations of the hierarchy, as
applying to the scattering within cell $i$:
 \be\label{phi1i}
 \hspace*{-.7cm}
 {\partial \langle \varphi_i({\bm k})
 \rangle_Y\over {\partial Y}}= {\bar\alpha_s}\!\int
 \frac{d^2{\bm p}}{2\pi}\, {{{\bm k}^2} \over {{\bm p}^2 ({\bm k}-{\bm p})^2}}
 \Big\langle \,2\,{{\bm p}^2 \over {\bm k}^2}\, \varphi_i({\bm p}) -
 \,\varphi_i({\bm k})\Big\rangle_Y -
 {\bar\alpha_s}\,
 \big\langle \varphi^2_i({\bm k})\big\rangle_Y,
 \nn
 \ee
and, respectively,
 \be\label{phi2i}
 {\partial \langle \varphi_i(\bm{k}_1)\varphi_i(\bm{k}_2)
 \rangle_Y\over {\partial Y}}&=&{\bar\alpha_s}\Big\{\!\int
 \frac{d^2{\bm p}}{2\pi}
\, {{{\bm k}_1^2} \over {{\bm p}^2 ({\bm k}_1-{\bm p})^2}}
 \Big\langle \,\Big(2\,{{\bm p}^2 \over {\bm k}_1^2}\,
 \varphi_i({\bm p}) -
 \,\varphi_i({\bm k}_1)\Big)
 \varphi_i(\bm{k}_2)\Big\rangle_Y
 \nn&{}&\qquad -\,
 \big\langle \varphi^2_i({\bm k}_1)\varphi_i(\bm{k}_2)\big\rangle_Y,
 \nn&{}&\qquad +\,\kappa\, \alpha_s^2\,
 \delta^{(2)}(\bm{k}_1-\bm{k}_2)\,
 \bm{k}_1^2\,\lan\varphi_i(\bm{k}_1)\ran_Y\Big\}
 \nn&{}&\qquad +\,\big\{1 \leftrightarrow 2\big\}\,.
 \ee
Clearly, these equations are (quasi)local in ${\bm b}$ in the sense
that they involve only operators pertaining to cell $i$. Needless to
say, this property holds also for the higher equations in the
hierarchy. Within the present, quasi--homogeneous, approximation,
different cells in the impact parameter space evolve independently
from each other.

It is now straightforward to identify the Langevin equation which
generates the same correlations as the hierarchy at cell $i$. This
reads (from now on, we omit the cell index $i$, as there is no
coupling between different cells):
 \be\label{LanQCD}
 \frac{1}{\bar\alpha_s}\, {\partial  \varphi({\bm k})
 \over {\partial Y}}&=&\!\int
 \frac{d^2{\bm p}}{2\pi}\, {{{\bm k}^2} \over {{\bm p}^2 ({\bm k}-{\bm p})^2}}
 \Big ( \,2\,{{\bm p}^2 \over {\bm k}^2}\, \varphi({\bm p}) -
 \,\varphi({\bm k})\Big) -  \varphi^2({\bm k})
 \nn&{}&\qquad +\,\sqrt{
 2{\kappa}\,\alpha_s^2 \,
 \varphi({\bm k})}\,\nu({\bm k})
 \ee
where $\nu({\bm k}, Y)$ is a Gaussian white noise : $\lan \nu({\bm
k}, Y)\ran=0$ and
 \be
 \lan \nu({\bm k}_1,Y_1)\nu({\bm k}_2,Y_2)\ran
 \,=\,\frac{1}{\bar\alpha_s}\,
 \delta(Y_1-Y_2)\,\delta^{(2)}(\bm{k}_1-\bm{k}_2)\,
 \bm{k}_1^2\,.
 \ee
Eq.~(\ref{LanQCD}) must be understood with a rapidity discretization
prescription of the Ito type, as explicitly shown in Eq.~(\ref{toylangdiscr}).

Eq.~(\ref{LanQCD}) should be compared to Eq.~(\ref{Balitsky}) which,
we recall, is the generating equation of the Balitsky hierarchy.
Clearly, Eq.~(\ref{Balitsky}) is the same as the (Fourier transform
of the) deterministic part of Eq.~(\ref{LanQCD}), but the latter
involves also the noise term responsible for fluctuations. In that
respect, Eq.~(\ref{LanQCD}) is closer to the Langevin equation
(\ref{toylang}) for the statistical toy model. Note, however, that
in contrast to Eq.~(\ref{toylang}), the noise term in the
corresponding equation (\ref{LanQCD}) for QCD does not vanish at
saturation. This reflects the fact that, in our previous analysis of
QCD, we did not include the effect of fluctuations on the
recombination terms (an effect which would go beyond the dipole
picture that we have used to describe fluctuations in QCD).
However, this omission is truly harmless: In the high density regime
at/near saturation, fluctuations are anyway unimportant, and there
is no need to suppress them explicitly\footnote{Numerical
simulations within the context of statistical physics
\cite{PL99,Moro042} have confirmed that the sFKPP equation, in which
the noise term has as a coefficient
$\sqrt{\alpha\varphi(1-\varphi)}$ --- with $\alpha\ll 1$ and
$\varphi=1$ at saturation --- and the so--called Reggeon model,
where the coefficient of the noise is simply $\sqrt{\alpha\varphi}$,
lead indeed to identical results for the measured correlations.}. In
fact, it is rather obvious on Eq.~(\ref{LanQCD}) that the effects of
the noise term are important only so long as $\varphi({\bm k})\simle
\alpha_s^2$.

\section{Physical discussion}\label{Physics}
\setcounter{equation}{0}

The Langevin equation (\ref{LanQCD}) turns out to be the natural
generalization to QCD of the stochastic
Fisher--Kolmogorov--Petrovsky--Piscounov (sFKPP) equation (see the
Appendix), to which it reduces in a standard approximation to the
BFKL kernel known as the `diffusion approximation'. This will be
discussed in Sect. 7.1. Then, in Sects. 7.2 and 7.3, we shall use
some known results about the sFKPP equation \cite{PL99,Moro042,MS95}
(see also Refs. \cite{Saar,Panja} for recent reviews and more
references) in order to explore the physical consequences of
Eq.~(\ref{LanQCD}) for QCD. In particular, in Sect. 7.2, we shall
make the connection with the physical discussion and the results in
Ref. \cite{IMM04}. Then, in Sect. 7.3, we shall discover a rather
dramatic consequence of the fluctuations, namely the breakdown of
the BFKL approximation in the regime where the average gluon density
is small.

\subsection{Relation with the sFKPP equation}

In order to understand the relation between our Langevin equation
Eq.~(\ref{LanQCD}) in QCD and the sFKPP equation in statistical
physics, it is useful to first recall a few facts about the
deterministic part of Eq.~(\ref{LanQCD}), as obtained after
neglecting the noise term there. If we also neglect correlations in
the initial conditions, then this deterministic part is simply the
momentum--space version of the BK equation, which can be more
compactly rewritten as:
 \be\label{BKshort} 
 \partial_Y  \varphi(\rho,Y)\,=\,{\bar\alpha_s}\,
  \chi(-\partial_\rho)\varphi\,-
  {\bar\alpha_s}\,\varphi^2\,\ee
where $\rho\equiv\ln k^2/k^2_0$ (with $k_0$ some arbitrary momentum
scale of reference),
$\chi(\gamma)=2\psi(1)-\psi(\gamma)-\psi(1\!-\!\gamma)$ is the
Mellin transform of the BFKL kernel, and  $\chi(-\partial_\rho)$ is
an integro--differential operator defined via the series expansion
of $\chi(\gamma)$ (see below). In a series of papers \cite{MP03},
Munier and Peschanski have argued that Eq.~(\ref{BKshort}) is in the
same universality class as the
Fisher--Kolmogorov--Petrovsky--Piscounov (FKPP) equation, which
appears as a mean field approximation to a variety of stochastic
problems in chemistry, physics, and biology \cite{Saar,FKPP}. To
understand this correspondence, notice that in the dilute regime at
high transverse momenta, $k^2\gg Q_s^2(Y)$ or $\rho\gg \rho_s(Y)$,
with $ \rho_s(Y)\equiv\ln Q_s^2(Y)/k^2_0$, we have $\phi\ll 1$, so
the dominant behavior of $\varphi(\rho,Y)$ is determined by the
linear part of Eq.~(\ref{BKshort}), which is the BFKL equation
\cite{BFKL}. One thus finds that the dominant dependencies upon
$\rho$ and $Y$ can be isolated out into an exponential factor:
 \be
 \varphi(\rho,Y)\,=\, e^{-\gamma_0(\rho-\lambda_0\bar\alpha_s Y)}
 \,\psi(\rho,Y), \label{front}\ee
where $\gamma_0$ and $\lambda_0$ are pure numbers determined by the
BFKL kernel as \cite{GLR} :
 \be\label{gamma0} \gamma_0 \chi'(\gamma_0)=\chi(\gamma_0),\qquad
 \lambda_0= \frac{\chi(\gamma_0)}{\gamma_0},\ee
which implies $\gamma_0\approx 0.63$ and $\lambda_0=\chi'(\gamma_0)
\approx 4.88$, and the function $\psi(\rho,Y)$ is comparatively
slowly varying, so its behavior can be studied by using a limited
expansion of the operator $\chi(-\partial_\rho)$ around
$\chi(\gamma_0)$ :
 \be\label{kernelexp}
 \chi(-\partial_\rho)&=& \chi(\gamma_0) +
 \chi'(\gamma_0)(-\partial_\rho-\gamma_0)+\frac{1}{2}
 \chi''(\gamma_0)(-\partial_\rho-\gamma_0)^2\,\dots\nn
 &\approx& -\lambda_0\partial_\rho + D_0
 (-\partial_\rho-\gamma_0)^2,\ee
where $D_0\equiv \chi''(\gamma_0)/2$ and in writing the second line
we have also used Eq.~(\ref{gamma0}). The approximation which
consists in keeping only the terms to second order in this
expansion, as explicitly shown in the second line above, is
generally referred to as the ``diffusion approximation", and is
equivalent to a saddle point approximation to the solution to BFKL
equation in Mellin space.
v
In the case of BFKL equation, one can use Eqs.~(\ref{front}) and
(\ref{kernelexp}) to show that $\psi(\rho,Y)$ obeys a diffusion
equation. For the non--linear BK equation (\ref{BKshort}), one can
argue that the main effect of the non--linear term is to introduce
an absorptive boundary condition on the diffusion equation for
$\psi$ \cite{MT02}. Alternatively, one can study the full
non--linear equation which emerges within the diffusion
approximation:
  \be\label{BKexp} 
 \partial_Y  \varphi(\rho,Y)\,=\,{\bar\alpha_s}\,\big(
  -\lambda_0\partial_\rho + D_0
  (-\partial_\rho-\gamma_0)^2\big)\varphi \,-
  {\bar\alpha_s}\,\varphi^2\,.\ee
Munier and Peschanski have observed that, up to a linear change of
variables and an appropriate rescaling of $\varphi$,
Eq.~(\ref{BKexp}) is the same as the FKPP equation:
 \be\label{KPP}
 \partial_t u(x,t)= \partial_x^2u(x,t) + u(x,t)\big(1-u(x,t)\big),
 \ee
which at large times describes an uniformly translating front which
propagates from the stable state $u=1$ into the unstable one $u=0$
and decays exponentially at large $x$ (far ahead the front)
\cite{Saar,FKPP}. In QCD, this front is already visible on
Eq.~(\ref{front}), which at large $Y$ describes a traveling wave
located at $\rho =\rho_s(Y)\approx \lambda_0\bar\alpha_s Y$ which
has an exponential slope $\gamma_0$ and propagates in `time'
$\bar\alpha_s Y$ with uniform velocity $\lambda_0$. (This behavior
is the origin of geometric scaling \cite{geometric} for the BK
equation \cite{SCALING,MT02}.) By using known properties of the FKPP
equation \cite{Saar} or, alternatively, by solving the diffusion
equation for $\psi$ with an absorptive boundary condition
\cite{MT02}, one can deduce the dominant corrections to the velocity
and the shape of the front due to non--linearities. For large $Y$
and $\rho\gg \rho_s(Y)$, one finds $\psi(\rho,Y)\sim (\rho
-\rho_s(Y))/Y^{3/2}$, and therefore:
\begin{equation}
\frac{d\rho_s(Y)}{dY}\,\simeq\, \bar\alpha \lambda_0
-\frac{3}{2\gamma_0}\frac{1}{Y}\ . \label{satscal0}
\end{equation}
Returning to the original Langevin equation (\ref{LanQCD}) and
assuming that the dominant $\rho$--behavior of the solution is still
given by the exponential $e^{-\gamma_0\rho}$ (which turns out to be
right indeed), one can use again  the diffusion approximation
(\ref{kernelexp}) to deduce a simplified form of the equation
($\tau\equiv \bar\alpha_s Y$):
 \be\label{LANexp} \hspace*{-.5cm}
 \partial_\tau \varphi(\rho,\tau)\,=\,
  -\lambda_0\partial_\rho \varphi + D_0
  (-\partial_\rho-\gamma_0)^2\varphi \,-
  \,\varphi^2\,+\,\sqrt{2{\kappa\, \alpha_s^2}
  \,\varphi}\,\nu(\rho,\tau),\ee
with the Gaussian white noise :
  \be
  \lan \nu(\rho,\tau)\ran=0,\qquad \lan \nu(\rho,\tau)\nu(\rho',\tau')\ran
  = \,\frac{1}{\pi}\,\delta(\tau-\tau')\,
  \delta(\rho-\rho')\,.
  \ee
Up to a simple change of variables, the equation above is
essentially the same as the sFKPP equation \cite{Panja}:
 \be\label{sFKPP}\hspace*{-.7cm}
 \partial_t u(x,t)= \partial_x^2u \,+\, u(1-u)
 \,+\,\sqrt{\frac{2}{N}\,u(1-u)}\,\nu(x,t)\,.
 \ee
In this analogy, the density of particles per site $u(x,t)$ in the
stochastic particle model corresponds to the scattering amplitude
(or unintegrated gluon distribution) $\varphi(\rho,\tau)$ in QCD,
and $1/\alpha_s^2$ plays the same role as $N$ (the number of
particles per site at saturation), which is reasonable since we
recall that $n\sim 1/\alpha_s^2$ is also the dipole occupation
number at saturation. Therefore, the mean field approximation
emerges in the limit where the occupation numbers are large, i.e.,
$N\to\infty$ for the particle model and $\alpha_s^2\to 0$ in QCD.

Note finally the limitations of this correspondence, as inherent
in the use of the diffusion approximation. First, Eq.~(\ref{LANexp})
predicts that behind the front $\varphi$ saturates at a constant
value $\varphi=\gamma_0^2 D_0$ (or $u=1$ for Eq.~(\ref{sFKPP})),
whereas the original Langevin rather yields $\varphi({k},Y)
\approx \ln (Q_s(Y)/k)$ [i.e.,
$\varphi(\rho,Y)\approx (\rho_s(Y)-\rho)/2\,$]
for $\rho \ll \rho_s(Y)$, as most easily seen by using
Eq.~(\ref{phiN}) together with the
fact that $T(r)=1$ for $r \gg 1/Q_s(Y)$. Second, for fixed $Y$
and sufficiently large $\rho$ (essentially, such that $\rho-
\rho_s(Y)\simge \rho_s(Y)$ \cite{SCALING}), the BFKL
`anomalous dimension' $1-\gamma$ approaches to zero, which signals
the transition to a regime dominated by the DGLAP dynamics. Clearly,
in this regime it is not possible to expand the BFKL kernel around
the saturation exponent $\gamma_0$, as we did in Eq.~(\ref{kernelexp}).
Thus, we do not expect the simplified equation (\ref{LANexp})
to describe correctly the transition to the DGLAP regime.

\subsection{Some results from sFKPP equation and their consequences
for QCD}

Even if somehow simpler than the original Lagevin equation
(\ref{LanQCD}), the sFKPP equation (\ref{LANexp}) (or (\ref{sFKPP}))
remains complicated, because of the simultaneous presence of the
non--linear term for recombination and of the multiplicative noise
term. Fortunately, this equation has been extensively studied in
relation with problems in statistical physics
--- mostly through numerical simulations, but also via some
analytical methods ---, with results that we shall briefly describe
here and then adapt to the QCD problem at hand. In applying these
results to QCD, one should however keep in mind the possible
limitations of the correspondence between Eqs.~(\ref{LanQCD}) and
(\ref{sFKPP}), as mentioned at the end of the previous subsection.

It has been rigorously demonstrated \cite{MS95} that the front
generated by the sFKPP equation (\ref{sFKPP}) is {\it compact} : For
any $t$, there exists a $x_r(t)$ such that $u(x,t)= 0$ for $x >
x_r(t)$, and also a $x_l(t)$ such that $u(x,t)=1$ for $x< x_l(t)$.
One expects a similar property for Eq.~(\ref{LanQCD}), although in
that case the transition to the saturation region behind the front
may not be as sharp, because the coefficient of the corresponding
noise term does not vanish at saturation.

Numerical simulations show that for large but finite $N$, the front
generated by Eq.~(\ref{sFKPP}) propagates with an asymptotic
velocity $v_N$ which is smaller than the corresponding velocity
$v_0=2$ for the FKPP equation (\ref{KPP}). Moreover, with increasing
$N$ the convergence of $v_N$ towards $v_0$ is extremely slow:
$v_0-v_N\sim 1/\ln^2 N$ when $N\gg 1$. For the corresponding QCD
problem, this implies the following asymptotic $Y$-dependence of the
saturation scale :
\begin{equation}
\lambda_s\equiv\,\lim_{\tau\to\infty}\,
\frac{d\rho_s(\tau)}{d\tau}\,\simeq\,\lambda_0\,-\,
\frac{\cal C}{\ln^2(1/\alpha_s^2)}\qquad {\rm when}\quad \alpha_s\ll 1,
\label{satscal}
\end{equation}
(recall that $\tau = \bar\alpha_s Y$). Thus, with decreasing $\alpha_s$,
the {\it saturation exponent} $\lambda_s$
converges only slowly towards the respective mean--field value in
Eq.~(\ref{gamma0}).

Since the noise term in Eq.~(\ref{LanQCD}) is important only for relatively
small $\varphi\simle \alpha_s^2$, we expect the shape of an individual
front (in its comoving frame) to be rather well described by the solution
(\ref{front}) to the deterministic equation (\ref{BKshort})
at all the points where $\varphi\gg \alpha_s^2$ (or $u\gg 1/N$ for
Eq.~(\ref{sFKPP})). This is seen indeed in the numerical
simulations. One can then estimate the width of the front as
follows: It is the noise term in Eq.~(\ref{LanQCD}) which abruptly
cuts down the growth of $\varphi$ ahead of the front, but this
requires $\varphi$ to be as small as $\alpha_s^2$. Since in the
front region behind the tip $\varphi$ behaves like $\varphi\simeq
e^{-\gamma_0(\rho-\rho_s)}$, cf. Eq.~(\ref{front}), we conclude that
$\varphi$ decreases from $1$ to $\alpha_s^2$ over a range
$\rho-\rho_s\sim (1/\gamma_0)\ln (1/\alpha_s^2)$, which should be a
good estimate for the width of the front.

However, because of the fluctuations inherent in the noise term,
different realizations of the same evolution will lead to an {\it
ensemble} of fronts which all have the same shape, but are displaced
with respect to each other along the $\rho$--axis. That is, the
position $\rho_s$ of the front is itself a random variable,
characterized by an expectation value $\lan\rho_s(\tau)\ran$, which
for large $\tau$ increases according to Eq.~(\ref{satscal}) (since
this is the common asymptotic behavior of all the fronts in the
ensemble), and also by a dispersion $\sigma^2\equiv \lan\rho_s^2\ran
- \langle\rho_s\rangle^2$, which is expected to rise linearly with
$\tau$ : $\sigma^2(\tau) \sim D_{\rm fr}\tau$ (since the front
executes a random walk around its average position). The numerical
simulations to the sFKPP equation confirm this behavior, and show
that the {\it front diffusion coefficient} $D_{\rm fr}$ scales like
$1/\ln^3 N$ when $N\gg 1$. For QCD, this in turn implies:
\be\label{Dfr} D_{\rm fr}\,\simeq\,\frac{\cal
D}{\ln^3(1/\alpha_s^2)} \qquad {\rm when}\quad \alpha_s\ll 1\,,\ee
which vanishes, as expected, when $\alpha_s\to 0$, but only very
slowly.

Note that the results (\ref{satscal}) and (\ref{Dfr}) are only
logarithmically sensitive to the coefficient of ${\cal
O}(\alpha_s^2)$ of the noise term in Eq.~(\ref{LanQCD}), or
(\ref{LANexp}). Thus, the leading order estimates in the limit
$\alpha_s\to 0$, as shown in Eqs.~(\ref{satscal}) and (\ref{Dfr}),
are not affected by our uncertainty concerning the fudge factor
$\kappa$ in the noise. On the other hand, the next--to--leading
order correction already will be sensitive to the precise
coefficient under the log, and thus to $\kappa$ (since, e.g.,
$1/\ln^2(a/\kappa\alpha_s^2) \approx 1/\ln^2(1/\alpha_s^2) -
2\ln(a/\kappa)/\ln^3(1/\alpha_s^2)$).

The results of the sFKPP equation alluded to above turn out to be
consistent with numerical and analytic studies of the stochastic
particle models whose mean field approximation is the FKPP equation
(\ref{KPP}). This agreement should not come as a surprise: As shown
in the Appendix, the sFKPP equation emerges precisely as the
continuum description of such discrete particle models. In
particular, in that context, Brunet and Derrida \cite{BD} have given
a simple heuristic argument which explains the $1/\ln^2 N$ scaling
of the velocity correction $v_0-v_N$ at large $N$, and also allows
one to compute the corresponding coefficient ${\cal C}$. Namely,
they have observed that, in the presence of discreteness, diffusion
should replace local growth as the main mechanism for front
propagation: Indeed, in order for the growth term (the linear term
in $u(x,t)$ in Eq.~(\ref{KPP})) to be effective, there must be at
least one particle per bin (or lattice site). Thus, the only way
that a particle can move to an originally empty bin ahead of the
front is via diffusion from the previously occupied bins on its
left. To mimic that, Brunet and Derrida proposed a modified
deterministic equation obtained by inserting a cutoff
$\theta(u-1/N)$ in the growth term in Eq.~(\ref{KPP}) (since a
density $u\sim 1/N$ corresponds to a site occupation number of
${\cal O}(1)$). A simple analysis of this equation then implies
\cite{BD} $v_0-v_N\simeq {\cal C}/\ln^2 N$ for $N\gg 1$, with a
value for ${\cal C}$ which is indeed consistent with the numerical
studies of both particle models and the sFKPP equation. After
translation to the QCD problem of interest here, this in turn
implies:
 \be\label{ls}
 \lambda_s\,\simeq\,\lambda_0\,-\,
 \frac{\pi^2 \gamma_0 \chi''(\gamma_0)}{2\ln^2 (1/\alpha_s^2)}
 \,\qquad {\rm when}\quad \alpha_s\ll 1
 .\ee
It so happens that the QCD coefficient ${\cal C} = \pi^2 \gamma_0
\chi''(\gamma_0)/2$ is numerically large, ${\cal C}\approx 150$, so
the corrective term in the equation above can be trusted only for
extremely small values of $\alpha_s$, which are physically
unrealistic. To our knowledge, there is no analytic argument
allowing one to understand the scaling (\ref{Dfr}) of the front
diffusion coefficient, or to compute the coefficient ${\cal D}$
there. (See however the discussion in Sect. 4 of Ref. \cite{Panja}.)

Furthermore, in the context of QCD, the above results
(\ref{satscal}) and (\ref{Dfr}) of the sFKPP equation corroborate
the conclusions obtained in Ref. \cite{IMM04} through an
analogy between the high--energy problem in QCD
and some specific particle models in statistical physics.
This in turn demonstrates that the Langevin equation (\ref{LanQCD}),
or, more generally, the stochastic equations presented in
Sect. \ref{NewEqs}, provide the correct evolution law
underlying the physical picture put forward in Ref. \cite{IMM04}.
By (numerically) solving these equations, one can now go beyond the
analysis in Ref. \cite{IMM04} and study the evolution for realistic
(non--asymptotic) values of $Y$ and $\alpha_s^2$.

\subsection{Front diffusion and the breakdown of the BFKL approximation}

It has been argued in Ref. \cite{IMM04} that, as a consequence of the
diffusive wandering of the front, the `geometric scaling' property
characteristic of the individual fronts --- i.e.,
the fact that a particular
front realization propagates as a travelling wave,
$T(\rho,Y)\approx T(\rho-\rho_s(Y))$ (cf. Eq.~(\ref{front})),
so that its shape in the comoving frame is not changed under
the evolution --- is actually broken after averaging over the
statistical ensemble of fronts generated by the stochastic
evolution up to large $Y$.
In what follows, we shall demonstrate that the consequences
of the front diffusion are in fact even more dramatic, as they entail
the breakdown of the BFKL approximation in the dilute regime,
where this approximation is usually assumed to work\footnote{We
are grateful to Al Mueller for helping us clarifying this point.}.
What we mean by that more precisely is that, at very high energies,
the standard BFKL equation (i.e., the
linearized part of Eqs.~(\ref{EQT1}) or (\ref{BKshort})) fails to
correctly describe the evolution of the average scattering amplitude
$\lan T(\rho)\ran_Y$ even in the regime where this amplitude is small,
$\lan T(\rho)\ran_Y\ll 1$. 
This is so because, in the presence of fluctuations and for
sufficiently large $Y$, average quantities like $\lan T\ran_Y$ or
$\lan \Tt\ran_Y$ are dominated by those fronts within the
statistical ensemble which are at saturation for the values of
$\rho$ of interest, and this even when $\rho$ is well above the
average saturation momentum $\lan\rho_s\ran_Y$, where $\lan
T(\rho)\ran_Y\ll 1$ indeed.

Note that in this subsection we return to the notation $T$ (rather
than $\varphi$) for the amplitude, that is, we prefer to work in
coordinate space, where $\rho=\ln (1/r^2k_0^2)$. Then, for a given
front realization, with saturation momentum $\rho_s(Y)$, the
scattering amplitude saturates to $T=1$ behind the front, and takes
the scaling form $T(\rho,Y)\simeq e^{-\gamma_0(\rho-\rho_s)}$
 within a finite range in $z$ ahead of the front.
(Recall that the front is compact.) For the present purposes, it
suffices to describe this behavior with the simple interpolation:
\begin{equation}\label{Tevent}
    T(\rho,\rho_s)=
    \begin{cases}
        \displaystyle{1} &
        \text{ for\,  $\rho \leq \rho_s$}
        \\
        \displaystyle{\exp \left[ -\gamma_0 (\rho - \rho_s) \right]} &
        \text{ for\,  $\rho \geq \rho_s$},
    \end{cases}
\end{equation}
which neglects the compact nature of the front; this is harmless
since, as we shall see, the tail of the distribution at large $z$
does not contribute to average quantities in the regime of interest.

Because of the diffusive nature of the front wandering, the values
of $\rho_s$ are distributed according to the probability density
\begin{equation}\label{probdens}
    P(\rho_s) =
    \frac{1}{\sqrt{\pi}\sigma}\,
    \exp \left[
    -\frac{\left( \rho_s - \langle \rho_s \rangle \right)^2}{\sigma^2}
    \right],
\end{equation}
where $\langle \rho_s \rangle\simeq \lambda_s\bar\alpha_s Y$ and
$\sigma^2\simeq D_{\rm fr}\bar\alpha_s Y$, as discussed before. Then
the average amplitude $\lan T \ran$ is determined by
\begin{equation}
    \lan T(\rho,\lan \rho_s \ran) \ran =
    \int \limits_{-\infty}^{\infty}
    d\rho_s\, P(\rho_s)\, T(\rho,\rho_s),
\end{equation}
and higher--point correlations can be computed similarly. For
simplicity, in what follows we shall restrict ourselves to higher
correlations evaluated at equal points, e.g., $\lan
\Tt(\rho,\rho)\ran\equiv \lan T^2(\rho) \ran$. Then, one can obtain
$\lan T^n \ran$ by simply replacing $\gamma_0 \rightarrow n\gamma_0$ in
the subsequent formulae for $\lan T\ran$.

By using Eqs.(\ref{Tevent}) and Eq.(\ref{probdens}) it is
straightforward to show that
\begin{equation}\label{Tave}
    \lan T \ran=
    \frac{1}{2}\, {\rm Erfc}\left(\frac{z}{\sigma} \right)
    +\frac{1}{2}
    \exp\left( \frac{\gamma_0^2 \sigma^2}{4} - \gamma_0 z\right)
    \left[ 2 - {\rm Erfc}\left(\frac{z}{\sigma} -
    \frac{\gamma_0 \sigma}{2} \right)\right],
\end{equation}
where $z\equiv \rho - \lan \rho_s
\ran $ and ${\rm Erfc(x)}$ is the complimentary error function, for
which we recall that
\begin{equation}\label{erfc}
    {\rm Erfc}(x)=
    \begin{cases}
        \displaystyle{2-\frac{\exp(-x^2)}{\sqrt{\pi}x}} &
        \text{ for\,  $x \ll -1$}
        \\*[0.1cm]
        \displaystyle{1} &
        \text{ for\,  $x=0$}
        \\*[0.1cm]
        \displaystyle{\frac{\exp(-x^2)}{\sqrt{\pi}x}} &
        \text{ for\,  $x \gg 1$}.
    \end{cases}
\end{equation}
For what follows it is important to notice that the two terms in the
r.h.s. of Eq.(\ref{Tave}) arise from the saturating piece and,
respectively, the exponentially decaying piece, of
Eq.(\ref{Tevent}).

The behavior of $\lan T \ran$ as a function of $z$ depends upon the
competition between $\sigma$ (the width of the Gaussian distribution
of the fronts) and $1/\gamma_0$, which characterizes the exponential
decay of the individual fronts. Since $\sigma$ grows like $\sqrt{Y}$
whereas $\gamma_0 \sim {\cal O}(1)$, one may conclude that the
typical situation at high energy is such that $\sigma\gg
1/\gamma_0$. But the diffusion coefficient (\ref{Dfr}) vanishes when
$\alpha_s\to 0$, even though only slowly, so for sufficiently small
$\alpha_s$ at fixed $Y$ one can imagine also the situation where
$\sigma\ll 1/\gamma_0$. So, we shall consider both cases here:

\noindent $\bullet$ \underline{$\sigma \ll 1/\gamma_0$}

In this regime, the front diffusion should play no role, and this is
indeed what we find. Specifically, when $z \ll -\sigma$, by making
use of Eq.(\ref{erfc}) one sees that the first term in
Eq.(\ref{Tave}) gives 1, while the second one is negligible. On the
contrary, when $z \gg \sigma$, the first term is small while the
second one gives $\exp(-\gamma_0 z)$. Thus, not surprisingly, one
finds that the average amplitude $\lan T \ran$ retains the single
event profile, except in the short interval $|z| \lesssim \sigma$
where it gets smoothed. The BFKL dynamics, characterized by the
anomalous dimension $\gamma_0$, is still visible, and the mean field
approximation holds since $\lan T^n \ran \simeq \lan T \ran^n$.

\noindent $\bullet$ \underline{$\sigma \gg 1/\gamma_0$}

In this regime, the correlations are dominated by fluctuations, and
the BFKL behavior is washed out everywhere except at extremely large
distances ahead of the front (where however the present
approximations cannot be trusted since, e.g., the compact support
property of the front, and also higher correlations in the diffusion
of the  front, may play a role).

Specifically, for all values of $z$ such that $z \ll
\gamma_0\sigma^2$ one finds that $\lan T \ran$ is dominated by the
first term in Eq.(\ref{Tave}):
\begin{equation}\label{Thighsigma}
    \lan T \ran \,\simeq\,
    \frac{1}{2}\, {\rm Erfc}\left(\frac{z}{\sigma} \right)
    \qquad {\rm for} \quad -\infty < z \ll \gamma_0\sigma^2,
\end{equation}
which, as mentioned earlier, is the contribution from the saturating
pieces of the single events, and thus is independent of $\gamma_0$.
This estimate holds, in particular, in the range $\sigma \ll z \ll
\gamma_0\sigma^2$ where $\lan T \ran$ is small, $\lan T \ran\ll 1$,
yet very different from the corresponding BFKL prediction. In
particular, there is no trace of geometric scaling, 
in agreement with Refs. \cite{MS04,IMM04}.

In fact, within the whole range in $z$ in which
Eq.(\ref{Thighsigma}) is valid, the higher correlations $\lan T^n
\ran$ are given by this same expression, that is
\begin{equation}\label{TnequalTn}
    \lan T^n \ran \simeq
    \lan T \ran
    \qquad {\rm for} \quad -\infty < z \ll \gamma_0\sigma^2,
\end{equation}
which signals a total breakdown of the mean field approximation,
except in the saturation regime where $\lan T \ran\simeq 1$. One can
even find a window within which $\lan T^n \ran$ varies very slowly,
namely:
\begin{equation}\label{Tnequalhalf}
    \lan T^n \ran \simeq
    \lan T \ran \simeq \frac{1}{2}
    \qquad {\rm for} \quad  |z| \ll \sigma.
\end{equation}

\begin{figure}[t]
    \centerline{\epsfxsize=7.cm\epsfbox{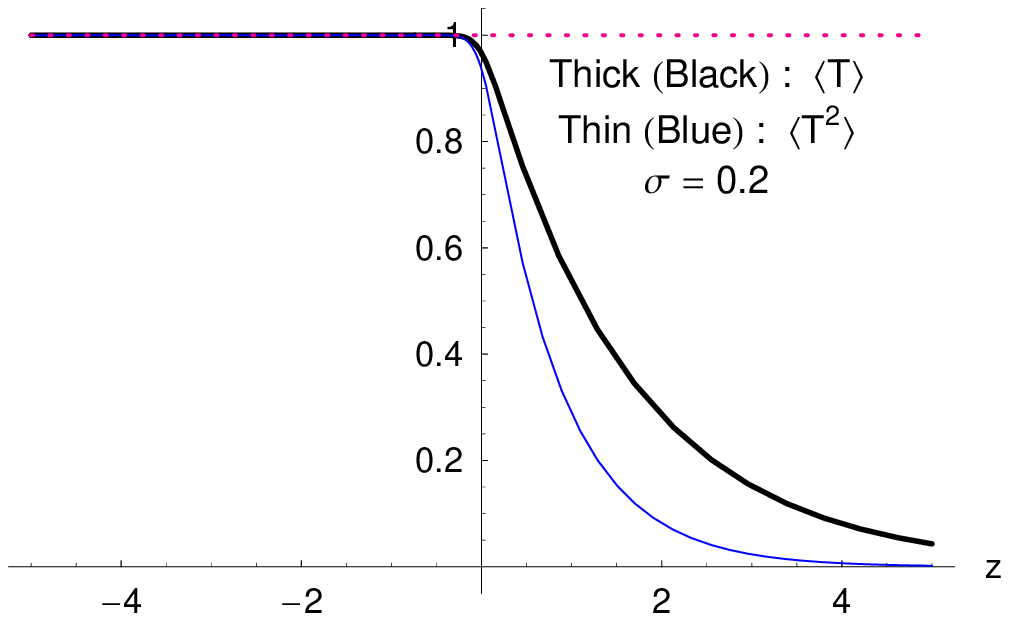}
    \hspace{0.25cm}
    \epsfxsize=7.cm\epsfbox{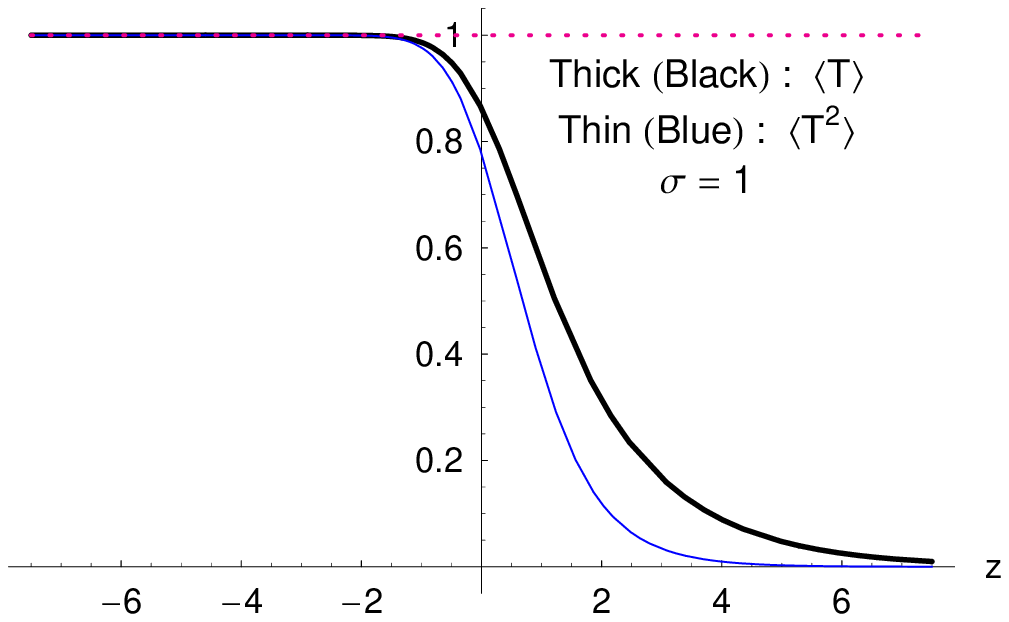}}
    \vspace{0.75cm}
    \centerline{\epsfxsize=7.cm\epsfbox{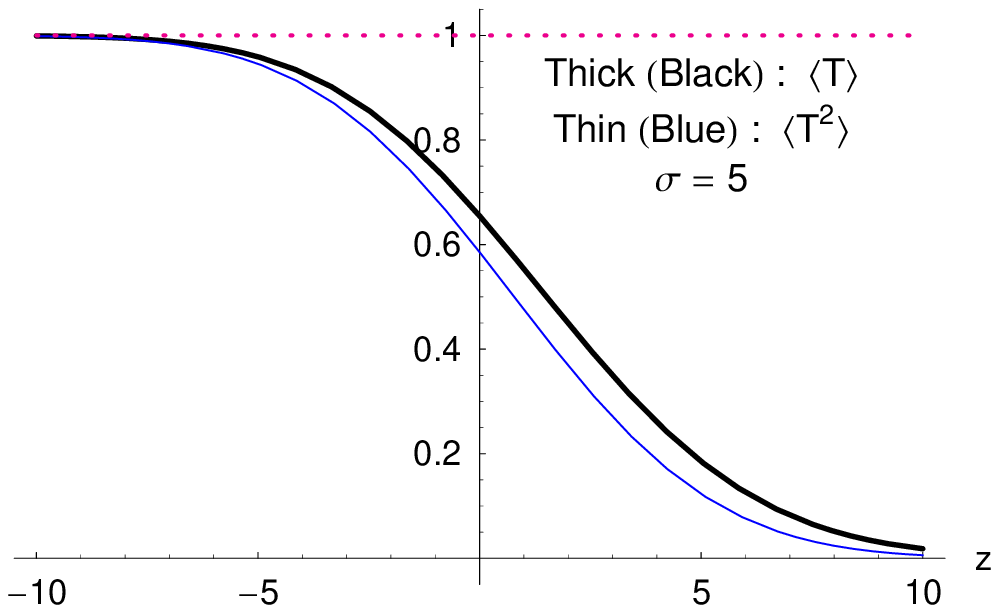}
    \hspace{0.25cm}
    \epsfxsize=7.cm\epsfbox{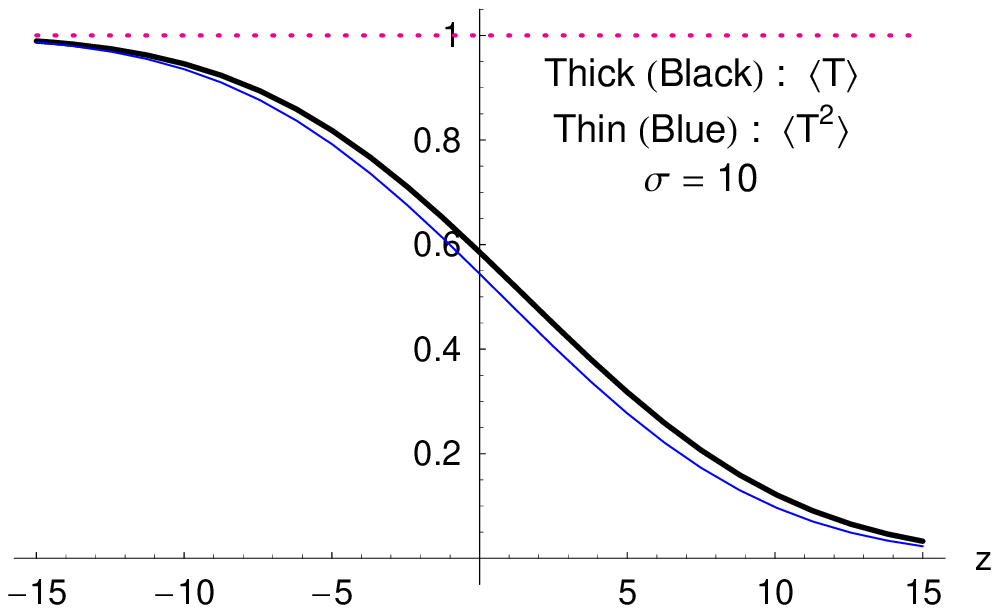}}
    \caption{\label{sigma}\sl Evolution of $\langle T \rangle$ and
                 $\langle T^{2} \rangle$ with increasing $\sigma$.}
\end{figure}

The different behaviors encountered when increasing $\sigma$ (and
thus $Y$) are illustrated on the example of $\lan T \ran$ and $\lan
T^2 \ran$ in Fig.~\ref{sigma}.

Let us conclude this section, and also the paper, with a final
remark concerning the role of fluctuations in the high--energy
evolution in QCD, and, in particular, on their interplay with
saturation. It has been a recurrent theme in this paper that
fluctuations in the dilute regime act as a seed for the growth of
higher--point correlations which then, through their subsequent
evolution, generate the non--linear terms responsible for
saturation. This is the (more or less) expected part of the
scenario, in which fluctuations and saturation are important at
different ends of the spectrum, and the amplitude $T(\rho,Y)$ is
reasonably well described by the mean field approximation (the BK
equation) everywhere except at the tip of the distribution where $T$
is very small: $T\simle \alpha_s^2$. This picture is indeed what
emerges from the dynamical equation (\ref{LanQCD}), but {\it only so
long as one considers a single front realization} (i.e., a single
event). If, on the other hand, one considers the {\it statistical
ensemble of fronts}
--- which is what one needs to do in order to compute average
quantities ---, then the dispersion of the fronts due to
fluctuations has a rather unexpected consequence, which is to render
the correlation functions sensitive to saturation even in the
(formally) weak scattering regime, where the {\it average}
scattering amplitude is small, and thus wash out the normal BFKL
behavior.

\section*{Acknowledgments}

The ideas presented in this paper have crystallized over a rather
lengthy period of time, during which we have benefited from
illuminating discussions and useful suggestions from several
colleagues. In particular, the efforts developed by one of us
(E.I.) together with Kazu Itakura and Al Mueller towards
understanding `pomeron loops' in the context of non--linear
evolution have certainly had an important influence on the
gestation of the work presented here.
We are particularly grateful to Larry McLerran and Al Mueller
for helping us clarifying several important points in the present
analysis. We are equally grateful to Jean-Paul Blaizot, Fran\c cois
Gelis, Kazu Itakura, Robi Peschanski and Gavin Salam
for their comments on this manuscript and other insightful remarks.
We further acknowledge useful conversations with Ian Balitsky,
Yoshitaka Hatta, Yuri Kovchegov, Misha Lublinsky, Stephane Munier,
Arif Shoshi, Raju Venugopalan, Samuel Wallon, and Stephen Wong.

\appendix

\section{A one--dimensional reaction-diffusion model}
In this Appendix we shall study a much more sophisticated model,
than the one we did in Sect.~\ref{Toy}, which shares a lot of similar
features with QCD. We consider a system whose state is described by
the number of particles A at every site $i$ in a one-dimensional
lattice; each state is of the form $\{n_i\}=\{n_1,n_2,...,n_i,...\}$,
where $n_i$ is the number of particles at site $i$. The dynamics
contains three possible processes:

\noindent $\bullet$ A particle can split locally into two at a rate
$\alpha$, i.e.
\begin{equation}
        {\rm A}_i \xrightarrow[]{\alpha} {\rm A}_i + {\rm A}_i.
\end{equation}
$\bullet$ Two particles can recombine locally into one at a rate $2
\beta_0$, i.e.
\begin{equation}
        {\rm A}_i + {\rm A}_i \xrightarrow[]{2 \beta_0} {\rm A}_i.
\end{equation}
$\bullet$ A particle can diffuse to a neighboring site at a rate
$\mu$, i.e.
\begin{equation}
        {\rm A}_i \xrightarrow[]{\mu} {\rm A}_{i+1}, \qquad
        {\rm A}_i \xrightarrow[]{\mu} {\rm A}_{i-1}.
    \end{equation}
Our task is to derive the evolution equation for $\lan n_i \ran$ and
$\lan n_i n_j \ran$ from the master equation, which we do not write
explicitly here, but the corresponding changes in probabilities will
be obvious from the analysis below. Then we will study the
continuous limit, which will lead us to the necessity of requiring
normal-ordering, and we will end this Appendix by introducing the,
relevant to the problem, Langevin equation.

Since splitting and recombination take place locally, their
contribution to the time evolution of $\lan n_i \ran$ is easy to
obtain. For both these processes, one can have either gain or loss
under a step $dY$, depending on the number of particles in the initial
configuration. There are two terms arising from the splitting part
$dP(\{n_i\};Y)/dY|_{\alpha}$ (all the other splittings which occur
at lattice points different than the one we measure, i.e.~the $i$-th
one, cancel each other):\\

\noindent $\circ$  Loss $\qquad -\alpha \, n_i \,P(...,n_i,...;Y)$

\noindent $\circ$ Gain $\qquad +\alpha \, (n_i-1)
\,P(...,n_i-1,...;Y),$\\

\noindent and two terms from the recombination part
$dP(\{n_i\};Y)/dY|_{\beta_0}$\\

\noindent $\circ$ Loss $\qquad -\beta_0 \, n_i\, (n_i-1)
\,P(...,n_i,...;Y)$

\noindent $\circ$ Gain $\qquad+\beta_0 \, (n_i+1)\,n_i
\,P(...,n_i+1,...;Y)$.\\

\noindent To calculate $d \lan n_i \ran /dY|_{\alpha,\beta_0}$, we
appropriately shift the $n$'s wherever needed to obtain
\begin{equation}
    \frac{d \lan n_i \ran}{dY}\bigg|_{\alpha,\beta_0}=
    \sum_{\{n_i\}} P(\{n_i\};Y)\, F_{\alpha,\beta_0}(\{n_i\}),
\end{equation}
where, after very simple algebra, one can find that
$F_{\alpha,\beta_0}(\{n_i\})=\alpha\, n_i - \beta_0 (n_i^2-n_i)$.
Therefore we arrive at
\begin{equation}\label{1dimnab}
    \frac{d \lan n_i \ran}{dY}\bigg|_{\alpha,\beta_0}=
    \alpha \lan n_i \ran -
    \beta_0 \left[
    \lan n_i^2 \ran -
    \lan n_i \ran
    \right].
\end{equation}
Now let us also study in detail the evolution of $\lan n_i \ran$ due
to diffusion. Under a step $dY$ in time, the diffusion can drive an
particle from site $i$ to $i\pm1$. Again, in all cases we can have
either gain or loss depending on the number of particles in the
initial configuration. Overall there are eight terms arising from
the diffusive part $dP(\{n_i\};Y)/dY|_{\mu}$:

\noindent $\circ$ Loss
\begin{align}
    &-\mu \, n_i \, P(...,n_{i-1},n_i,n_{i+1},...;Y)
    & {\rm from} \quad i \rightarrow i-1
    \nonumber \\
    &-\mu \, n_i \, P(...,n_{i-1},n_i,n_{i+1},...;Y)
    & {\rm from} \quad i \rightarrow i+1
    \nonumber \\
    &-\mu \, n_{i-1} \, P(...,n_{i-1},n_i,n_{i+1},...;Y)
    & {\rm from} \quad i-1 \rightarrow i
    \nonumber \\
    &-\mu \, n_{i+1} \, P(...,n_{i-1},n_i,n_{i+1},...;Y)
    & {\rm from} \quad i+1 \rightarrow i
    \nonumber
\end{align}
\noindent $\circ$ Gain
\begin{align}
    &+\mu \, (n_i+1) \, P(...,n_{i-1}-1,n_i+1,n_{i+1},...;Y)
    & {\rm from} \quad i \rightarrow i-1
    \nonumber \\
    &+\mu \, (n_i+1) \, P(...,n_{i-1},n_i+1,n_{i+1}-1,...;Y)
    & {\rm from} \quad i \rightarrow i+1
    \nonumber \\
    &+\mu \, (n_{i-1}+1) \, P(...,n_{i-1}+1,n_i-1,n_{i+1},...;Y)
    & {\rm from} \quad i-1 \rightarrow i
    \nonumber \\
    &+\mu \, (n_{i+1}+1) \, P(...,n_{i-1},n_i-1,n_{i+1}+1,...;Y)
    & {\rm from} \quad i+1 \rightarrow i
    \nonumber
\end{align}
As before, in order to calculate $d \lan n_i \ran /dY|_{\mu}$, we
appropriately shift the $n$'s wherever needed to obtain
\begin{equation}
    \frac{d \lan n_i \ran}{dY}\bigg|_{\mu}=
    \sum_{\{n_i\}} P(\{n_i\};Y)\, F_{\mu}(\{n_i\}),
\end{equation}
with $F_{\mu}(\{n_i\})=\mu (n_{i+1}+n_{i-1}-2 n_i)$. Therefore we
arrive at
\begin{equation}\label{1dimnmu}
    \frac{d \lan n_i \ran}{dY}\bigg|_{\mu}=
    \mu \lan \nabla^2 n_i \ran,
\end{equation}
where, for reasons that will become obvious shortly, we introduced
the shorthand notation
\begin{equation}\label{1dimdel}
    \nabla^2 n_i \equiv
    n_{i+1}+ n_{i-1}-2 n_{i+1}.
\end{equation}
Now, when considering the pair density we need to study the three
separate cases $\lan n_i^2 \ran$, $\lan n_i n_{i+1} \ran$ and $\lan
n_i n_j \ran$ for $|i-j| \geq 2$. Here we shall skip the derivation,
which can be done by following the same steps of the $\lan n_i \ran$
case, and only give the final set of the evolution equations which
read
\begin{align}
    \frac {d \lan n_i n_j \ran}{dY} \bigg|_{\alpha}= &\,
    \alpha \lan
    2 n_i n_j +\delta_{ij} n_i
    \ran,
    \\
    \frac {d \lan n_i n_j \ran}{dY} \bigg|_{\beta_0}= &\,
    -\beta_0 \lan
    n_i^2 n_j + n_i n_j^2 -2 n_i n_j
    -\delta_{ij} (n_i^2 - n_i)
    \ran,
    \\ \label{1dimn2mu}
    \frac {d \lan n_i n_j \ran}{dY} \bigg|_{\mu}= &\,
    \mu \big \langle
    n_i \nabla^2 n_j + n_i \nabla^2 n_j
    \nonumber \\
    & +\delta_{ij} (\nabla^2 n_i + 4 n_i)
    -(\delta_{i,j-1} + \delta_{i,j+1}) (n_i +n_j)
    \big \rangle.
\end{align}

Let's turn our attention to the continuous limit. Let the lattice
spacing be equal to $\Delta$, the coordinate $x=i\Delta$ and the
particle single and pair densities $\bar{n}(x)=n_i/\Delta$ and
$\bar{n}(x_1) \bar{n}(x_2)=n_i n_j/\Delta^2$. In order to be
economical in the notation we shall drop the bars immediately. Then
the single density equation becomes
\begin{align}
    \frac{\del \lan n(x) \ran}{\del Y}=
    \alpha \lan n(x) \ran -
    \beta_0 \Delta \left[
    \lan n^2(x) \ran -
    \frac{1}{\Delta}\lan n(x) \ran
    \right]
    + \mu \Delta^2\, \frac{\del^2 \lan n(x) \ran}{\del x^2}
    +\mathcal{O}(\mu \Delta^4).
\end{align}
Notice that the next to last term justifies the shorthand notation
adopted in Eq.(\ref{1dimdel}). Considering first this diffusion
term, we naturally impose that $\mu \Delta^2 \equiv D = {\rm fixed}$
in the $\Delta \rightarrow 0$ limit. Then all the higher order terms
can be ignored. Similarly, we require $\beta_0 \Delta \equiv \beta =
{\rm fixed}$ for the recombination term. It is only the
recombination process which generates potential singularities in the
continuous limit, since it involves the pair density. However, even
though the second term in the square bracket becomes divergent, we
expect the first term to do so in the same limit in such a way to
cancel the singularity. Indeed, the whole bracket corresponds to the
$x_1 \rightarrow x_2$ limit of the average normal-ordered pair
density
\begin{equation}\label{1dimnormal2}
    n^{(2)}(x_1,x_2)
    \equiv \lan :\!n(x_1) n(x_2)\!:\ran=
    \lan n(x_1) n(x_2) \ran - \delta(x_1-x_2) \lan n(x_1) \ran,
\end{equation}
and thus we finally arrive at
\begin{align}\label{1dimconn}
    \frac{\del n^{(1)}(x)}{\del Y}=
    \alpha  n^{(1)}(x)
    -\beta n^{(2)}(x,x)
    + D \,\frac{\del^2 n^{(1)}(x)}{\del x^2},
\end{align}
with the obvious identification $n^{(1)}(x)\equiv \lan n(x)\ran$. Of
course, in the absence of recombination, these subtleties would have
not appeared. But they do appear for all the processes, i.e. even
for splitting and diffusion, when we try to take the continuous
limit in the equation of the pair density. As an illustration, let's
concentrate on the diffusion contribution. When we consider the
$|i-j| \geq 2$ case we arrive at an equation with no singularities,
since all the terms in the second line of Eq.(\ref{1dimn2mu})
vanish. But those linear terms do not vanish when we start from the
discrete equation with $|i-j| \leq 1$ (while the quadratic terms do
not change). In fact, they are divergent in the limit $\Delta
\rightarrow 0$. Moreover they are ``scheme-dependent'', in the sense
that, even though all cases $|i-j| \leq 1$ lead to an equation for
$\lan n^2(x) \ran$, this equation is not uniquely obtained; it
depends on the particular discrete equation that one starts with. We
will not write all the details here, but we shall give the most
divergent terms. We easily find that
\begin{equation}
    \frac{\del \lan n^2(x,x) \ran}{\del Y}\bigg|_{\rm div}=
    \begin{cases}
        \displaystyle{
        4\, \frac{D}{\Delta^3}\,
        \lan n(x) \ran
        +\mathcal{O}\left( \frac{1}{\Delta^2}\right)} &
        \text{ from \, $i=j$}
        \\*[0.5cm]
        \displaystyle{
        -2\, \frac{D}{\Delta^3}\,
        \lan n(x) \ran
        +\mathcal{O}\left( \frac{1}{\Delta^2}\right)} &
        \text{ from \, $i=j \pm 1$}.
    \end{cases}
\end{equation}
From the previous discussion, we expect that the divergencies will
disappear when we consider the evolution of normal ordered
densities. Indeed, using Eq.(\ref{1dimnormal2}) and
\begin{align}\label{1dimnormal3}
    n^{(3)}(x_1,x_2,x_3) = &\,
    \lan n(x_1) n(x_2) n(x_3) \ran
    -\left[
    \delta(x_1-x_2) \lan n(x_1) n(x_3)\ran
    + {\rm perm.} \right]
    \nonumber \\
    &+2\, \delta(x_1-x_2) \delta(x_1-x_3) \lan n(x_1) \ran,
\end{align}
and along with Eq.(\ref{1dimconn}) we obtain a unique and finite
evolution equation of the pair density in the elegant form
\begin{align}\label{1dimconn2}
    \frac{\del n^{(2)}(x_1,x_2)}{\del Y}=&\,
    \bigg\{
    \alpha \left[
    n^{(2)}(x_1,x_2) +\delta(x_1-x_2) \,n^{(1)}(x_1)
    \right]
    \nonumber \\
    &-\beta \left[n^{(3)}(x_1,x_1,x_2)
    +\delta(x_1-x_2)\, n^{(2)}(x_1,x_1) \right]
    \nonumber\\
    &+ D n^{(1)}(x_2)\, \frac{\del^2 n^{(1)}(x_1)}{\del x_1^2}
    \bigg\} + \left\{1 \leftrightarrow 2 \right\}.
\end{align}
In principle one could follow the same procedure to derive the
evolution equation for the general $\kappa$-point function
$n^{(\kappa)}(x_1,...,x_{\kappa})$ or understand the form by looking
at the first two equations, (\ref{1dimconn}) and (\ref{1dimconn2}),
which we just derived. We shall not write the full hierarchy here,
but the easiest and most rewarding way to derive it is from the
corresponding Langevin equation. This is simply the generalization
of the Langevin equation we saw in Sec.~\ref{Toy} and reads
\begin{equation}\label{1dimlang}
    \frac{\del \tilde{n}(x)}{\del Y} = \alpha\, \tilde{n}(x)
    -\beta\, \tilde{n}^2(x)
    +D\, \frac{\del^2 \tilde{n}(x)}{\del x^2}
    +\sqrt{2 \left[ \alpha\, \tilde{n}(x)
    -\beta\, \tilde{n}^2(x) \right]}\,\, \nu(x,Y),
\end{equation}
where the noise satisfies $\lan \nu(x,Y) \ran = 0$ and $\lan
\nu(x_1,Y) \nu(x_2,Y') \ran = \delta(x_1-x_2) \delta(Y-Y')$. Notice
that $\tilde{n}(x)$ saturates at the value $\alpha/\beta$. The
hierarchy generated by Eq.(\ref{1dimlang}) is equivalent to the one
of the model we studied, provided we simply identify $\lan
\tilde{n}(x_1)...\tilde{n}(x_{\kappa}) \ran$ with
$n^{(\kappa)}(x_1,...,x_{\kappa})$. One sees that in a
straightforward way by deriving Eqs.(\ref{1dimconn}) and
(\ref{1dimconn2}) from Eq.(\ref{1dimlang}), while a formal proof can
be found in \cite{PL99}.

For our convenience, and in order to make a closer analogy to the
QCD problem (where $T$ saturates to one),
we can make the change of variables $\tilde{n}
\rightarrow (\alpha/\beta) u$, $Y \rightarrow t/\alpha$, $D
\rightarrow \alpha D$ and $\nu \rightarrow \sqrt{\alpha}\, \nu$.
Then the Langevin equation becomes
\begin{equation}\label{ulang}
    \frac{\del u(x)}{\del t} =
    u(x) -u^2(x) + D\, \frac{\del^2 u(x)}{\del x^2}
    +\sqrt{\frac{2\beta}{\alpha}
    \left[u(x)-u^2(x) \right]}\, \,\nu(x,t),
\end{equation}
with $\lan \nu(x,t) \ran = 0$ and $\lan \nu(x_1,t) \nu(x_2,t') \ran
= \delta(x_1-x_2) \delta(t-t')$. Now $u(x)$ saturates at the value
1.

\vfill

\end{document}